%% file: ms.tex
\documentclass[preprint]{aastex}

\begin{document}

\title{Absolute--Magnitude Distributions and Light Curves of
Stripped--Envelope Supernovae }

\author{Dean Richardson\altaffilmark{1,2}, David Branch\altaffilmark{1} and 
E.~Baron\altaffilmark{1}}

\altaffiltext{1}{Dept. of Physics and Astronomy, University of Oklahoma, Norman,
OK 73019}   
\altaffiltext{2}{Physics Dept., Marquette University, Milwaukee, WI 53201}    

\begin{abstract} 

The absolute visual magnitudes of three Type~IIb, 11~Type~Ib and
13~Type~Ic supernovae (collectively known as stripped--envelope
supernovae) are studied by collecting data on the apparent magnitude,
distance, and interstellar extinction of each event.  Weighted and
unweighted mean absolute magnitudes of the combined sample as well as
various subsets of the sample are reported.  The limited sample size
and the considerable uncertainties, especially those associated with
extinction in the host galaxies, prevent firm conclusions regarding
differences between the absolute magnitudes of supernovae of Type~Ib
and Ic, and regarding the existence of separate groups of overluminous
and normal-luminosity stripped--envelope supernovae.  The
spectroscopic characteristics of the events of the sample are
considered. Three of the four overluminous events are known to have
had unusual spectra.  Most but not all of the normal luminosity events
had typical spectra.

Light curves of stripped--envelope supernovae are collected and
compared.  Because SN~1994I in M51 was very well observed it often is
regarded as the prototypical Type~Ic supernova, but it has the
fastest light curve in the sample.  Light curves are modeled by means
of a simple analytical technique that, combined with a constraint on
$E/M$ from spectroscopy, yields internally consistent values of
ejected mass, kinetic energy, and nickel mass.  

\end{abstract} 


\section{Introduction}

In a recent paper \citep[][hereafter R02]{richardson02} we carried out
a comparative study of the absolute magnitudes of all supernovae (SNe)
in the Asiago Catalog.  Because of the large number of SNe in the
sample, we did not attempt to estimate the extinction of each SN in
its host galaxy, and we did not assign uncertainties to individual SN
absolute magnitudes.  In this paper we look more closely at the
absolute--magnitude distributions of stripped--envelope supernovae
(SE~SNe) by assigning uncertainties to each of the quantities that
enter into the absolute--magnitude determination, including host
galaxy extinction. By SE~SNe we mean SNe of Types IIb, Ib and Ic. (The
subset containing only SNe~Ib and Ic is referred to as SNe~Ibc.)  The
progenitors of SE~SNe are stars that have lost most or all of their
hydrogen envelopes. This can happen by strong winds such as in Wolf
Rayet stars or through mass transfer to a companion star such as in
Roche--lobe overflow or a common--envelope phase.  The light curves of
SE~SNe are powered by the radioactive decay of $^{56}$Ni, so the
absolute magnitudes are closely related to the ejected $^{56}$Ni
masses, and in turn to the stellar progenitors and explosion
mechanisms. Since the discovery of the apparent association of
GRB980425 with the peculiar Type~Ic SN~1998bw \citep{galama99}, and
the confirmation by spectra of a SN~1998bw--like event associated with
GRB030329 \citep{garnavich03}, SE~SNe have become of intense interest
in connection with GRBs. Contamination of high--redshift samples of 
Type~Ia supernovae by SE SNe also is an important issue \citep{homeier05}.  

We also study the $V$--band light curves (LCs) of SE~SNe.  Data were
collected from the literature for two SNe~IIb, seven SNe~Ib and 11
SNe~Ic including three that had unusually broad and blueshifted
absorption features in their spectra (we will refer to these as
hypernovae). The LCs show considerable diversity in peak brightness,
in the width of the peak, and the slope of the late--time tail. In
order to relate all of the LCs to total ejected mass, ejected nickel
mass, and kinetic energy in an internally consistent way, we fit the
data to a simple LC model.

The peak absolute--magnitude data and analysis are described in \S2.
The light curve data and model fits are presented in \S3.  A brief
summary is given in \S4.

\section{Absolute--Magnitude Distributions}

\subsection{Data}

R02 worked with the $B$ band, but for SE~SNe the $V$ band happens to
be the one for which most data are available.  In order to calculate
the peak visual absolute magnitude for each SN we collected data on
the peak apparent visual magnitude, the distance, the foreground
Galactic extinction, and the host galaxy extinction, all with assigned
uncertainties. We were able to find data for 27 events: three SNe~IIb
(Table~1), 11 SNe~Ib (Table~2) and 13 SNe~Ic (Table~3).  Eighteen
SNe~Ibc were in the sample of R02.

\placetable{\label{table1}}

\placetable{\label{table2}}

\placetable{\label{table3}}

\subsubsection{Peak Apparent Magnitudes} 

For most SNe the apparent magnitude and its uncertainty were taken
directly from the literature, but in some cases these values were not
given so it was necessary for us to estimate them. For SNe~1998dt,
1999di and 1999dn we used an uncalibrated $R$ band light curve
\citep{matheson01} together with a calibrated spectrum to determine
the peak $V$ magnitude. We used the spectrum that was nearest to
maximum light to calculate the $R$ magnitude at that epoch, and used
the $R$ light curve to determine the $R$ magnitude at peak.  Then we
calculated the $V-R$ color from the spectrum to determine $V$ at peak.
We examined the available data on $V-R$ versus time for SNe~Ibc and
estimated the total uncertainty in $V$ accordingly.  A similar method
was used for SN~1999cq \citep{matheson00}. The peak apparent magnitude
for SN~1992ar was taken from \cite{clocc00} who, from the limited data
available, presented two possible light curves, with different peak
magnitudes; we adopted the average.  In Tables~1 to 3 we can see that
the peak $V$ magnitudes are the dominant uncertainties in seven cases:
the SN~IIb 1987K, the SNe~Ib 1991D, 1998dt, 1999di and 1999dn, and 
the SNe~Ic 1992ar and 1999cq.   

\subsubsection{Distances}

Distance moduli were, for the most part, obtained as in R02.  When
possible we used a Cepheid calibrated distance to the host galaxy or a
galaxy in the same group as the host galaxy.  The second choice was
the distance given in Tully's Nearby Galaxies Catalog
\citep{tully88}, rescaled from $H_{0}$ = 75
km~s$^{-1}$~Mpc$^{-1}$ to our choice of $H_0$ = 60 for consistency
with R02.  We adopted an uncertainty of 0.2 magnitudes in the distance
modulus, combined in quadrature with the uncertainty resulting from
the radial velocity uncertainty of the host galaxy.  One significant
change since R02 is that a new distance, based on the tip of the red
giant branch, has become available for NGC~4214, the host of SN~1954A
\citep{drozdovsky02}. We adopt this distance in preference to the
(longer) Tully distance used in R02.  Now SN~1954A no longer appears to
be an overluminous SN~Ib.  Another change is the distance for
SN~1994I. This distance is taken from \cite{feldmeier97} who used the
Planetary Nebula Luminosity Function method.  In place of the Tully 
distance to 1990I, we use the distance given by \cite{elmhamdi04}
which was obtained by the same method.  The uncertainty was estimated
by considering the difference between the Tully distance and the
Elmhamdi distance.  The third choice was the luminosity distance
\citep{ski00} calculated from the redshift of the host galaxy (in each
of these cases $cz > 2000$ km s$^{-1}$) and assuming $H_0 = 60$,
$\Omega_M = 0.3$ and $\Omega_{\Lambda} = 0.7$.  A different choice of
$H_0$ would rescale the absolute magnitudes and different choices of
$\Omega_M$ and $\Omega_{\Lambda}$ would have very small effects on
this sample.  The uncertainty in the luminosity distance was
calculated assuming a peculiar velocity of 300 km~s$^{-1}$.  For many
of the events of our sample the uncertainty in the distance modulus is
significant, although it is dominant only in four cases: SNe~1954A,  
1990I, 1997ef and 1998bw.

\subsubsection{Extinction}  

The Galactic extinction is from \cite{sfd98}.  Values were taken from
NED\footnote{The NASA/IPAC Extragalactic Database (NED) is operated by
the Jet Propulsion Laboratory, California Institute of Technology,
under contract with the National Aeronautics and Space
Administration.} and converted from $A_B$ to $A_V$.  In all cases the
uncertainties in the Galactic extinction are comparatively small.

When possible the host galaxy extinction and its uncertainty were
taken from the literature.  When only the equivalent width of the
interstellar Na~I~D lines, $W(D)$, in the host was available we
calculated $E(B-V$) from the relation $E(B-V) = 0.16 W(D)$
\citep{turatto03} and then used $A_V = 3.1E(B-V)$. In this case we
took the uncertainty to be as large as the extinction, up to a maximum
uncertainty of 1.0 magnitude.

The host galaxy extinction for SN~1983V was taken from
\cite{clocc97}. \cite{porter87} reported that the HII region
associated with SN~1983V was not as prominent as the one associated
with SN~1962L but about the same as the one associated with SN~1964L.
Having only this information available we assigned the $A_V$(host)
value of SN~1983V to SN~1964L. For SN~1962L we took the average value
of all other SNe~Ic in our sample (which is larger than the extinction
of SN~1983V). For SNe~1962L and 1964L we assigned an uncertainty as
large as the extinction.

For SN~1990B we took the host galaxy extinction from
\cite{clocc01}. They quoted two values, one determined from the Na~I~D
line and one from the color excess; we chose to use the latter. They
did not quote an uncertainty except to say that it was large and
unknown, so we assigned a large uncertainty of 1.0 magnitude.

\cite{grothues91} give the extinction for various regions of NGC~3310,
the host galaxy of SN~1991N. Near the position of SN~1991N
\citep{barth96} the visual extinction was 1 to 2 magnitudes with an
uncertainty of about 1 magnitude. Because the SN was most likely
inside rather than behind the H~II region this extinction is probably
an overestimate. We adopted $A_V$ = 1.0 $\pm$ 1.0.

SE~SNe tend to be associated with star formation so they also tend to
be significantly extinguished in their host galaxies.  Tables~1 to 3
show that for many events in our sample the uncertainty in the host
galaxy extinction is the dominant uncertainty, and overall it is the
largest source of absolute--magnitude uncertainty for our sample.

\subsection{Analysis} 

\placefigure{\label{fig1}} 

Absolute visual magnitude is plotted against distance modulus in
Figure~1.  The slanted dashed line is the line of constant apparent
visual magnitude (less the total extinction) of 16. The horizontal
dashed line is the SN~Ia ``ridge line'' at $M_V=-19.5$, shown for
comparison.  The tendency of intrinsically brighter SNe to be at
larger distances, and the fact that all but a few of the SNe are to
the left of the slanted line, are obvious consequences of the strong
observational bias in favor of the discovery and follow--up of
brighter SNe.  The absence of SNe in the lower right part of the
figure is a selection effect against lower--luminosity events at large
distance, and the absence of SNe in the upper left is due to the fact
that overluminous events are uncommon, so none happen to have been
seen in relatively nearby galaxies.  Considering that the events of
this sample have been discovered in so many different ways, we make no
attempt to correct the absolute magnitude distribution for bias.
Instead, we emphasize that in this study we are simply characterizing
the available observational sample of SE~SNe; a sample in which
overluminous events are over--represented relative to less luminous
ones.

In Figure~1, there are quite a few SNe~Ic in the distance modulus 
range from about 29 to 32. There are also quite a few SNe~Ib in the 
distance modulus range from 33 to 35. These groups happen purely by 
chance. There is no spatial connection, other than distance, between 
the SNe within each of these groups.  

R02 considered the possibility that SNe~Ibc can be divided into two
luminosity groups: normal--luminosity SNe~Ibc, and overluminous
SNe~Ibc that are even more luminous than SNe~Ia.  We consider that
possibility here also.  As can be seen in Figure~1, four events of our
sample, three SNe~Ic and one SN~Ib, are above the SN~Ia ridge line.

The mean absolute magnitude and its standard deviation, both weighted
and unweighted, for the whole sample as well as for several subsets of
the sample, are given in Table~4.  The weighted mean of the whole
sample is $M_V=-18.03 \pm 0.06$, with $\sigma=0.89$.  When SE~SNe are
separated into normal--luminosity and overluminous we have $M_V=-17.77
\pm 0.06,\ \sigma=0.49$ for the normal--luminosity events and
$M_V=-20.08 \pm 0.18,\ \sigma=0.46$ for the overluminous.  Comparing
the normal--luminosity SNe~Ib and Ic, the unweighted means differ by
0.54 magnitudes in the sense that SNe~Ic are brighter than SNe~Ib, but
the weighted means differ by only 0.16 magnitudes so a difference
between normal--luminosity SNe~Ib and Ic is not firmly established by
these data.

\placetable{\label{table4}}
\placefigure{\label{fig2}} 

A histogram of the absolute magnitudes is shown in Fig.~2.  Fig.~2
also shows the best Gaussian fit to all of the data, determined by the
$\chi^2$ test using the mean absolute magnitude and dispersion as
parameters. The results were $\overline{M}_V$ = $-$18.49 and $\sigma$
= 1.13, but the low probability of 15\% confirms what is apparent to
the eye: the distribution is not adequately fit by a Gaussian.

Considering the possibility of two luminosity groups we also fit the
data to a double--peaked distribution. To do this we used:

\begin{equation}
f(x)=f_0\biggl(w \exp\biggl[-\frac{(x-x_1)^2}{2\sigma_{1}^{2}}\biggr] +
\exp\biggl[-\frac{(x-x_2)^2}{2\sigma_{2}^{2}}\biggr]\biggr),
\end{equation}
 
\noindent with five parameters: $x_1$ and $x_2$ (the two mean absolute
magnitudes), $\sigma_1$ and $\sigma_2$ (the two dispersions) and the
weighting factor, $w$. The normalization factor, $f_0$, is equal to
$(1+w)^{-1}$.  The results for the double--peaked distribution are:
$\overline{M}_{V,1} = -20.31$, $\overline{M}_{V,2} = -18.20$,
$\sigma_1$ = 0.18, $\sigma_2$ = 0.81 and $w$ = 0.13.  The probability
of this fit is 39\%, still quite low.
  
\subsection{Comments on Spectra}

Here we briefly consider the extent to which SE~SNe that have normal
luminosities have typical spectra and SE~SNe that are overluminous
have peculiar spectra.

\subsubsection{Type~IIb}

At early times the spectra of SNe~II have conspicuous H$\alpha$ and
H$\beta$ P--Cygni features but at later times they resemble the
spectra of SNe~Ib because the Balmer lines are replaced by He~I lines.
The early Balmer lines in SN~1996cb are stronger and more similar to
those of SN~1987A than to those of SN~1993J \citep{qiu99}, which may
mean that SN~1996cb had a thicker hydrogen layer than SN~1993J.
Overall, however, the spectra of the three SN~IIb in the sample are
rather similar, and the peak absolute magnitudes are the same within
the uncertainties.

\subsubsection{Type~Ib}

\cite{branch02} studied the optical spectra of a dozen SNe~Ib selected
on the basis of having deep He~I absorption features.  The events of
that sample displayed a rather high degree of spectral homogeneity,
except that three also contained deep H$\alpha$ absorptions.  Of the
11 SNe~Ib in the present sample, seven were in the sample of
\cite{branch02}: SNe~1983N, 1984L, 1998dt, 1999dn, 1954A, 1999di, and
2000H, with the last three being the ``deep--H$\alpha$'' events.  We
find that all seven of these events have absolute magnitudes within
the normal SN~Ib range, and we see no significant difference between
the absolute magnitudes of the deep--H$\alpha$ events and the others.

The single available spectrum \citep{leibundgut90} of one of the
SNe~Ib in our present sample, SN~1984I, covers a limited wavelength
range so that little can be said except that it does appear to be a
SN~Ib.  Its absolute magnitude is within the normal range.

 The spectra of SN~1990I contained typical SN~Ib absorption features
but they were broader and more blueshifted than those of the Branch
et~al. (2002) sample (Elmhandi et~al. 2004), although not enough to
be considered a hypernova.  The absolute magnitude is within the normal
range.

SN~1991D has been discussed by \cite{benetti02} and \cite{branch03}.
Its He~I absorptions were less deep and the velocity at the
photosphere near the time of maximum light was lower than in the
events of the \cite{branch02} sample.  Thus the one overluminous SN~Ib
of our present sample also had an unusual spectrum.

SN~1999ex was observed by \cite{hamuy02} who referred to it as an
intermediate Type~Ib/c because of its relatively weak He~I lines.
\cite{branch03} refers to it as a ``shallow helium'' SN~Ib because its
He~I lines were clearly present, although weaker than in the events of
the \cite{branch02} sample.  While this event had an unusual spectrum,
according to Table~2 its absolute magnitude is within the normal
range.

To summarize SNe~Ib: the single overluminous SN~Ib of our sample had
an unusual spectrum, and most but not all (not SNe~1990I and 1999ex)
of the normal--luminosity events had normal spectra.

\subsubsection{Type~Ic}

Five events of our sample, SNe~1983I, 1983V, 1987M, 1990B, and 1994I,
can be said to have had typical SN~Ic spectra. The limited available
spectra of three others, SNe~1962L, 1964L, and 1991N, also show no
indication of peculiarity.  The absolute magnitudes of all eight of
these events are within the normal SN~Ic range.

As is well known, SN~1998bw was overluminous, and its absorption
features were very broad and blueshifted.  Two other SNe~Ic of our
sample, SNe~1997ef \citep{mazzali00} and 2002ap \citep{kinugasa02},
also had broad spectral features, although not as broad as those of
SN~1998bw; these two SNe~Ic were {\sl not} overluminous.

Apart from SN~1998bw, the other two overluminous SNe~Ic of our sample
are SNe~1992ar and 1999cq. \cite{clocc00} conclude that the one
available spectrum of SN~1992ar is remarkably similar to a spectrum of
the Type~Ic SN~1983V, which as mentioned above had typical spectra.
\cite{matheson00} interpret the one good spectrum of SN~1999cq as that
of a SN~Ic but with unusual (so far unique) narrow lines of He~I
superimposed.  The spectrum of SN~1999cq certainly is peculiar.

SN~1999as probably is the brightest SN~Ic known, with an absolute
magnitude brighter than $-21.4$ \citep{hatano01}, but since no peak
apparent magnitude is available it is not included in our present
sample. Its spectrum was quite unusual.

Summarizing SNe~Ic: two of the three overluminous events (or three of
four, counting SN~1999as) are known to have had unusual spectra.  Most
but not all (not SNe~1997ef and 2002ap) normal--luminosity events had
typical SN~Ic spectra.

\section{Light Curves} 

\subsection{Data} 

Light curve data in the $V$ band were found for most of the SNe in the
absolute--magnitude sample of \S2. For a few events only $R$--band or
unfiltered LC data were available. The LC data for many of the SNe
were available from the same reference as the peak magnitude. In some
cases we collected data from several sources in order to get as much
coverage as possible. For example, most of the data for SN~1994I were
taken from \cite{richmond96}, but two late--time data points were
added from \cite{clocc97}.  The SNe that have $V$--band LCs, with
references, are listed in Table~5.

\placetable{\label{table5}}
\placefigure{\label{fig3}} 
\placefigure{\label{fig4}} 

The LC data for SNe~IIb/Ib and Ic are shown in Figures~3 and 4,
respectively. The lines connect the symbols for each SN to help
distinguish the data of one SN from another but in some cases do not
depict the true shapes of the LCs.

The tails of SE~SNe are powered primarily by the deposition of
gamma--rays generated in the decay of $^{56}$Co.  Because gamma--rays
increasingly escape, the slopes of the LC are steeper than the
$^{56}$Co decay slope.  The only exception in the sample is SN~1984L
which has a slow late--time decay slope (Figure~3).

\placefigure{\label{fig5}} 
\placefigure{\label{fig6}} 

Figures~5 and 6 are like Figures 3 and 4, respectively, except for
covering a shorter time interval in order to show more detail around
the time of peak brightness.  The two SNe~IIb LCs shown here are very
similar and are less luminous than most SNe~Ibc.  Exceptional among
the SNe~Ib is the extremely luminous SN~1991D, which declined rapidly
after peak.  SN~1994I was very well observed and therefore often is
regarded as the prototypical SN~Ic, but it has the narrowest peak and
the fastest overall decline among the SNe~Ic of our sample.

Model fits for all of the LCs are shown in \S3.2 except for SNe~1954A,
1984I, 1991N, and 2000H because the coverage in their visual LCs was
too poor.

\subsection{The Model} 

Numerical light--curve calculations based on hydrodynamical explosion
models and various assumptions have been calculated by several groups
and compared to the LCs of selected SE SNe.  Here we take the approach
of adopting a simple analytical model and applying it to all of the SE
SNe in our sample.  This results in an internally consistent set of
explosion parameters (ejected mass, ejected nickel mass, and kinetic
energy) for all events.  The model is simple, but in view of the
evidence that SE SNe tend to be aspherical, most of the numerical
light--curve calculations also are oversimplified.

We use the model of \cite{arnett82} for the peak of the LC, and the
model of \cite{jeffery99} for
the tail.  The Arnett model applies at early times when the diffusion
approximation is valid and the Jeffery model applies at later times
when the deposition of gamma--rays dominates the LC.  As depicted in
Figure~7 our model LC switches abruptly from the the Arnett LC to the
Jeffery LC when the two LCs cross.  The underlying assumptions are
poorest at the time of the transition.  The basic assumptions are
spherical symmetry; homologous expansion; that $^{56}$Ni is centrally
concentrated rather than mixed outward in the ejecta;
radiation--pressure dominance at early times; constant optical opacity
at early times; and constant gamma--ray opacity at late times.  The
luminosity of the Arnett part is

\placefigure{\label{fig7}} 

\begin{equation} 
L_{A}(t) = \epsilon_{Ni} M_{Ni} \left(10^{-\frac{\zeta}{2.5}}\right) 
\left(e^{-x^2}\right) \int_{0}^{x} 2 z e^{(-2zy +z^2)} dz,  
\end{equation} 
where $x \equiv \frac{t}{\tau_m}$ and $y \equiv
\frac{\tau_m}{2t_{e,Ni}}$,

\begin{equation} 
\tau_m = \sqrt{\frac{\kappa_{opt}}{\beta c}
\sqrt{\frac{6 M_{ej}^3}{5 E_k}}},   
\end{equation} 

\begin{equation}
\epsilon_{Ni} = \frac{Q^{Ni}_{ph+PE}}{m_{Ni} t_{e,Ni}}
\end{equation} 
The luminosity of the Jeffery part is: 

\begin{equation} 
L_{J}(t) = \epsilon_{Ni} M_{Ni} \left\{ e^{-\frac{t}{t_{e,Ni}}} + G\left( 
e^{-\frac{t}{t_e,Co}} - e^{-\frac{t}{t_{e,Ni}}} \right)    
\left[ f_{PE}^{Co} + f_{ph}^{Co}\left( 1 - e^{-(\frac{t_0}{t})^2} \right) 
\right] \right\},  
\end{equation} 
where
\begin{equation} 
t_0 = \sqrt{\frac{M_{ej} \kappa_{\gamma}}{4 \pi v_a v_b}},   
\end{equation} 

\begin{equation} 
v_i = v_i^{93J} \sqrt{\frac{\left(\frac{E_k}{M_{ej}}\right)}
{\left(\frac{E_k}{M_{ej}}\right)_{93J}}},\ \ i = a,b, 
\end{equation} 

\begin{equation} 
G = \left(\frac{Q^{Co}_{ph+PE}}{Q^{Ni}_{ph+PE}}\right)
\left(\frac{m_{Ni}}{m_{Co}}\right)\left(\frac{t_{e,Ni}}{t_{e,Co} - t_{e,Ni}}  
\right) = 0.184641.  
\end{equation} 

The e--folding times for $^{56}$Co and $^{56}$Ni decay, $t_{e,Co}$ and
$t_{e,Ni}$, are 111 and 8.77 days, respectively.  The energy per decay, 
including energy from photons and electron--positron pairs but not
from neutrinos (which escape), are $Q_{ph+PE}^{Co} = 3.74$ MeV and
$Q_{ph+PE}^{Ni} = 1.73$ MeV.  For $^{56}$Co decay the fractions of
energy in photons $f_{ph}$ and in kinetic energy of positron $f_{PE}$
are 0.968 and 0.032, respectively.  The above quantities are from Table~1 of 
\cite{jeffery99}.  The optical and gamma--ray opacities are taken to be
$\kappa_{opt} = 0.4$ cm$^2$~g$^{-1}$ and $\kappa_{\gamma} = 0.04$
cm$^2$~g$^{-1}$ .  \cite{arnett82} defined $\beta$ as $4\pi(\alpha
I_M/3)$, where $\alpha I_M = 3.29$ for uniform density
\citep{arnett80}.

The model LC is bolometric. $V$--band LCs are generally regarded as
having similar shape, but in order to adjust the brightness of the
model to better match a $V$--band LC a correction $\zeta$ was used for
the Arnett part of the LC.  The value of $\zeta$ was determined by
calibrating the peak of the LC to that of a typical SN~Ia. This was
done by fixing $E_k$ to 1 foe (10$^{51}$ erg), $M_{ej}$ to 1.4
$M_{\odot}$, and $M_{Ni}$ to 0.6 $M_{\odot}$, then choosing
$\zeta=-1.48$ so that the peak absolute magnitude became $-$19.5.

The velocities $v_a$ and $v_b$ are the inner and outer velocities
within which the mass density can be regarded as roughly constant.
\cite{jeffery99} used velocities thought to be appropriate for
SN~1987A. A better approximation for SE~SNe is to use velocities
thought to be appropriate for SN~1993J (the most thoroughly observed
and modeled SE~SN) and to rescale these velocities with respect to
$E_k$/$M_{ej}$ for each SN (equation~7).  For SN~1993J we use $v_a =
1000$ km s$^{-1}$, $v_b = 10,000$ km s$^{-1}$, and
($E_k$/$M_{ej}$)$_{93J} =0.51$ foe/$M_{\odot}$, from
\cite{blinnikov98}.

We began our study with four adjustable model parameters: $E_k$,
$M_{ej}$, $M_{Ni}$, and t$_{shift}$, where t$_{shift}$ shifts the LC
on the time axis.  The best $\chi^2$ fit often was formally very good
but the model parameters were physically unreasonable.  To remedy this
problem we developed the following procedure for constraining the
$E_k$/$M_{ej}$ ratio using spectroscopic information.  We used the
parameterized supernova synthetic--spectrum code {\bf Synow}
\citep{branch02} to construct a relation between the wavelength of the
peak of an Fe~II blend near 5000~\AA\ and the {\bf Synow} input
parameter $v_{phot}$, the velocity at the photosphere (Figure~8).  We
measured the wavelength of the peak in spectra of each event and
obtained a value of $v_{phot}$ from Figure~8.  We defined $a(t) =
(E_k/M_{ej})/v_{phot}^2(t)$ and used values of $E_k/M_{ej}$ obtained
by others from numerical LC calculations for seven events of our sample
(references are in Table~6) to construct Figures~9 and 10, for
normal--luminosity SE~SNe and hypernovae, respectively.  In these
figures the dashed line is the adopted relation and the difference
between the dashed line and the solid lines is taken as the
uncertainty.  For the remaining events of the sample (those not having
$E_k/M_{ej}$ values from numerical LC calculations), we used
Figures~9 and 10 with our estimates of $v_{phot}(t)$ to obtain
estimates of $E_k$/$M_{ej}$.

\placefigure{\label{fig8}} 

\placefigure{\label{fig9}} 
\placefigure{\label{fig10}} 

With $E_k$/$M_{ej}$ thus estimated spectroscopically, and three rather
than four adjustable model parameters we obtained relatively good fits
with reasonable parameter values (in most cases).

\subsection{Results} 

The parameter values determined from the best model fits are listed in
Table~6. Because uncertainties are not available for each data point
in all LCs, we quote the uncertainty at peak, $\delta$$M_V$, as the
characteristic uncertainty.  The uncertainties in $M_V$ given in
Table~6 differ from those listed in Tables~1 -- 3 because here each
one has the uncertainty in $E_k$/$M_{ej}$ added in quadrature.

About half of the SNe modeled here have been modeled in similar studies; 
most by more sophisticated numerical models.
Since our model is an analytic model, it has the advantage of being fast 
and therefore, we can use it on a larger sample.  
Table~7 gives a comparison of our results to the results of the other studies. 
The values of $M_{ej}$ from these other models tend to be larger than ours 
by a factor of approximately two. 

\placetable{\label{table6}}
\placetable{\label{table7}}

Since at least some, if not all, hypernovae are associated with GRBs
there is likely some interaction between the GRB jet and the expanding 
SN shell. A first order approximation is to say that the 
SN occurs independent of the GRB. For our simple model this is a 
reasonable approximation.  

\subsubsection{Type IIb} 

SN~1993J is the only SE~SN that has been observed early enough to see
the break--out shock in the $V$--band LC.  Because the break--out
shock has not been incorporated into the model we are using, that part
of the LC has been omitted from our analysis.  The model fit for
SN~1993J (Figure~11a), with $E_k= 0.66$ foe and $M_{ej}=1.3~M_\odot$,
looks satisfactory although $\chi^2=2.14$ is somewhat large and probably
due to the small adopted value of $\delta$$M_V$.  Numerical LC
calculations were carried out by \cite{ybb95} and \cite{blinnikov98}:
the former imposed $E_k= 1$ foe and favored a model having
$M_{ej}=2.6~M_\odot$; the latter adopted a model having $E_k= 1.2$ foe
and obtained a fit with $M_{ej}=2.45~M_\odot$. If we impose 
E$_k = 1$ foe and let M$_{ej}$ vary, we get M$_{ej} = 1.6 M_{\odot}$
(Table~7).   

\placefigure{\label{fig11}} 

The model fit for SN~1996cb (Figure~11b), with $E_k=0.22$ foe and
$M_{ej}=0.9~M_\odot$, is slightly too dim at the peak but overall it
is satisfactory.

\subsubsection{Type Ib} 

Of the seven SNe~Ib LCs plotted in Figure~3, five are worth fitting.
The model fit to the fragmentary LC of SN~1983N (Figure~12a), with
$E_k=0.30$ foe and $M_{ej}=0.8~M_\odot$, is good.  
We were not able to obtain an acceptable fit to the entire LC of
SN~1984L (see figure~13b) because of the exceptionally slow decline in
the tail (Figure~3). However, when the data obtained later than 200
days after explosion were omitted, we did obtain a satisfactory fit 
(Figure~12c),
with $E_k=0.97$ foe and $M_{ej}=1.8~M_\odot$ (this result is denoted
``pk'' in Table~6). \cite{baron93} also had trouble fitting all of the 
data with detailed LC calculations, and were forced to what they regarded 
as an improbable model having a very small optical opacity and 
$E_k \simeq 20$ foe, M$_{ej} \simeq 50 M_{\odot}$.   

\placefigure{\label{fig12}} 

The LC of SN~1990I (Figure~12d) drops rapidly after 250 days and our
best fitting model, which has $E_k=0.67$ foe and $M_{ej}=1.2~M_\odot$,
cannot account for this.  The problem with the fit near peak
brightness is due to the compromise between trying to fit the tail and
the peak. If the data later than 250 days are ignored, we get a better
fit with only slight changes in the model parameters.

SN~1991D, the brightest SN in the sample, has an exceptional LC that
declines rapidly from the peak, yet from its spectra we obtain
$E_k$/$M_{ej}=0.13$ foe/$M_{\odot}$, the smallest value in the sample.
The model cannot reconcile these contradictory aspects (Figure~12e).
It is possible to get a good fit if we drop the constraint on the
$E_k$/$M_{ej}$ ratio, but then we obtain $E_k$/$M_{ej}=8$
foe/$M_{\odot}$ which is inconsistent with the spectra.
\cite{benetti02} used a semi--analytical model to fit the LC of
SN~1991D.  They suggested this peculiar event may have been a SN~Ia
exploding inside the extended helium--rich envelope of a companion
star.  If this is correct then although SN~1991D must be regarded as
Type~Ib according to SN spectral classification, physically it may
have more in common with SNe~Ia.

SN~1999ex has very good coverage around the peak but there are no data
for the tail. The model fit, with $E_k=0.30$ foe and
$M_{ej}=0.9~M_\odot$, is satisfactory.

Other studies have looked at the LCs of SN~1983N \citep{shigeyama90} and SN~1990I 
\citep{elmhamdi04} assuming E$_k = 1$  foe. Their results are listed 
in Table~7 along with our results for comparison; as well as what we  
found when we imposed E$_k = 1$ foe.  
In both cases our value for M$_{ej}$ was somewhat smaller and is increased when 
imposing E$_k = 1$ foe.  

\subsubsection{Type Ic}

Model fits were carried out for all of the SN~Ic LCs plotted in
Figure~4 except for SN~1991N. The fit for the limited LC of SN~1962L,
with $E_k=0.11$ foe and $M_{ej}=0.6~M_\odot$, is satisfactory
(Figure~13a).  $M_{Ni}=0.37 $ is more than half as high as
$M_{ej}$. The very low $\chi^2$ is due to the large uncertainty of
$\delta$$M_V =$ 0.85.

\placefigure{\label{fig13}} 

For SN~1983I (Figure~13b) there are no pre--peak data. Our fit, with
$E_k=0.33$ foe and $M_{ej}=0.7~M_\odot$, is satisfactory.
Our fit to the fragmentary LC of SN~1983V (Figure~13c) is satisfactory,
with $E_k=0.99$ foe and $M_{ej}=1.3~M_\odot$.
Our fit to the fragmentary LC of SN~1987M also is satisfactory
(Figure~13d), with $E_k=0.19$ foe and $M_{ej}=0.4~M_\odot$.

The fit for SN~1990B (Figure~13e), with $E_k=0.55$ foe and
$M_{ej}=0.9~M_\odot$ fits the date well except for the last data
point.  According to \cite{clocc01} that point is especially uncertain
because of the difficulty of subtracting the host galaxy light.
SN~1992ar is the brightest SN~Ic in the sample. The model fit
(Figure~13f), with $E_k=1.1$ foe, $M_{ej}=1.5~M_\odot$, and a high
value of $M_{Ni}=0.84 M_\odot$, is acceptable.

\placefigure{\label{fig14}} 

SN~1994I has a very narrow LC peak and the best model fit (Figure~14a), with
$E_k=0.55$ foe and $M_{ej}=0.5~M_\odot$, is too broad at peak. Thus
the value of t$_{rise}$ given in Table~6, which already is the
smallest value in the sample, is too large.

The last three LCs are those of three hypernovae.  As mentioned above,
these are SE~SNe that have very broad, blueshifted absorption features
at early times. The LC of SN~1997ef, which has a very broad peak and
appears to have a very late transition point (Figure~14b), is not well
fit by our model, which gives $E_k=3.3$ foe and
$M_{ej}=3.1~M_\odot$. 
The model fit for SN~1998bw (Figure~14c), with $E_k=31$ foe and
$M_{ej}=6.2~M_\odot$, is good except near the transition point.  This
is by far the highest value of $E_k$ for the events of the sample.
Figure~14d shows the fit for SN~2002ap, with $E_k =2.7$ foe and $M_{ej}
=$ 1.7 $M_{\odot}$. Overall it is not bad, although the model peak is
a bit dim and the model transition point is somewhat early. The model
has trouble finding a tail that fits both the transition point and the
two late time data points.  

Other studies have looked at the LCs of SN~1983I \citep{shigeyama90} 
and SN~1987M 
\citep{nomoto90} assuming E$_k = 1$  foe. Their results are listed
in Table~7 along with our results for comparison; as well as what we
found when we imposed E$_k = 1$ foe.
In both cases our value for M$_{ej}$ was somewhat smaller, but increased 
when imposing E$_k = 1$ foe. 

We compare our results for the three hypernovae with the results of other 
studies in Table~7. The value of E$_k$ given in Table~6 for the brightest 
of the three (SN~1998bw) is comparable to the values found by \cite{nakamura01} 
(E$_k = 50$ foe) and \cite{woosley99} (E$_k = 22$ foe).   
The dimmest of the three hypernovae (SN~1997ef) is compared to a study by 
\cite{iwamoto00} where they obtained E$_k = 8$ foe and M$_{ej} = 7.6 M_\odot$.  

\section{Summary} 

We have used the available data to characterize the
absolute--magnitude distributions of the SE~SNe in the current
observational sample.  Most SE~SNe have a ``normal'' luminosity, which
at $M_V=-17.77 \pm 0.06$ is about a magnitude and a half dimmer
than SNe~Ia.  One sixth of the current sample of SE~SNe are
overluminous, i.e., more luminous than SNe~Ia, but these are strongly
favored by observational selection so the true fraction of SE~SNe that
are overluminous is much lower than one sixth.  The small size of the
sample and the considerable absolute magnitude uncertainties,
especially those due to host galaxy extinction, still prevent an
absolute magnitude difference between SNe~Ib and Ic from being firmly
established.  Three of the four (or four of the five, counting
SN~1999as) overluminous SE~SNe are known to have had unusual spectra;
a few of the normal--luminosity SE~SNe also had unusual spectra.  Much
more data on SE~SNe are needed in order to better determine the
absolute--magnitude distributions, and to correlate absolute
magnitudes with spectroscopic characteristics.

Absolute light curves in the $V$ band (some fragmentary) are available
for two SNe~IIb, seven SNe~Ib, and 12 SNe~Ic including three
hypernovae.  Two of the SNe~Ib, SNe~1984L and 1991D, have LCs that are quite
different from those of the others.  The light curves of the SNe~Ic
are rather diverse.  The light curve of SN~1994I, often considered to
be a typical SN~Ic, actually is the most rapidly declining light curve
in the SN~Ic sample.  

The simple analytical light--curve model was applied to two SNe~IIb,
five SNe~Ib and 10 SNe~Ic.  Instead of assuming a kinetic energy, such
as the canonical one foe, an $E_k$/$M_{ej}$ ratio was estimated on the
basis of spectroscopy, and the model fits then produced internally
consistent values of $E_k$, $M_{ej}$, and $M_{Ni}$.

Reasonably good fits were obtained for the two SNe~IIb and three of
the five SNe~Ib.  The slowly decaying tail of the SN~1984L LC and the
rapid decline from the peak of the SN~1991D LC cannot be fit by the
model.  With the exception of the hypernova SN~1997ef, reasonable fits
were obtained for the SNe~Ic, with a considerable range in the
parameter values.  As expected, the hypernovae have high $E_k$ and
somewhat high $M_{ej}$, but only one of the three has high $M_{Ni}$.

Our values of $M_{ej}$ (and $E_k$) tend to be lower than those
obtained by others by means of numerical LC calculations, while our
values for $M_{Ni}$ are a little higher.  Some of the other numerical
calculations are based on an assumed canonical value of $E_k =1$
foe. Because our spectroscopic constraint on $E_k$/$M_{ej}$ together
with our LC model leads to lower $E_k$, and lower $E_k$ makes the LC
peak dimmer, we need slightly higher $M_{Ni}$ values to make the LC
peaks as bright as observed.

The diversity among SN~Ic LCs is of special interest in connection
with the ongoing search for SN signals in GRB afterglows.  In a
forthcoming paper we will apply the same modeling technique used in
this paper to the putative SN bumps in GRB afterglow light curves.
This will enable us to infer internally consistent SN parameter values
for comparison with the results of this paper, and to investigate the
issue of whether the GRBs and the associated SNe are coincident in
time or whether, as in the supranova model, the SN precedes the
gamma-ray burst.

\vspace{7mm}

We would like to thank Rollin C. Thomas for helpful comments and
suggestions. Support for this work was provided by NASA through grants
GO-09074 and GO-09405 from the Space Telescope Science Institute,
which is operated by the Association of Universities for Research in
Astronomy, Inc., under NASA contract NAS 5-25255.  Additional support
was provided by National Science Foundation grants AST-9986965 and
AST-0204771.

\clearpage

\input{figures}

\clearpage 

\begin{deluxetable}{lccccccc}
\tabletypesize{\scriptsize}
\tablecaption{Absolute--Magnitude Data for SNe~IIb \label{table1}}
\tablewidth{0pt}
\tablehead{
\colhead{SN} &
\colhead{Galaxy} &
\colhead{V} &
\colhead{$\mu$} &
\colhead{$A_V$(Galactic)} &
\colhead{$A_V$(host)} &
\colhead{$M_V$} &
}
\startdata
1987K & NGC4651 & 14.4 $\pm$ 0.3(1)  & 31.09 $\pm$ 0.03$^a$(2)
& 0.088 $\pm$ 0.014 & 0.2 $\pm$ 0.2(3) & -17.0 $\pm$ 0.4  \\
1993J & NGC3031 & 10.86 $\pm$ 0.02(4)  & 27.80 $\pm$ 0.08$^a$(5)
& 0.266 $\pm$ 0.043 & 0.36 $\pm$ 0.22(6) & -17.57 $\pm$ 0.24  \\
1996cb & NGC4651 & 13.90 $\pm$ 0.03(7) & 31.09 $\pm$ 0.03$^a$(2)
& 0.100 $\pm$ 0.016 & 0.10 $\pm$ 0.10(7) & -17.39 $\pm$ 0.11 \\
\enddata
\tablenotetext{a} {Cepheid calibrated distance}
\tablenotetext{b} {Nearby Galaxies Catalog \citep{tully88}}
\tablenotetext{c} {$Luminosity\ Distance$ ($H_0$=60, $\Omega_M$=0.3,
$\Omega_{\Lambda}$=0.7; references are for redshifts)}    
\tablerefs{ 
(1) \cite{filippenko88}, (2) Average of Cepheid distances for NGC4321,
NGC4535, NGC4548 and NGC4639 in the same group; \cite{freedman01},
(3) \cite{filippenko87}, (4) \cite{vandriel93}, (5) \cite{freedman01},
(6) \cite{richmond94}, (7) \cite{qiu99}
}
\end{deluxetable} 

\clearpage 

\begin{deluxetable}{lccccccc}
\tabletypesize{\scriptsize}
\tablecaption{Absolute--Magnitude Data for SNe~Ib \label{table2}}
\tablewidth{0pt}
\tablehead{
\colhead{SN} &
\colhead{Galaxy} &
\colhead{V} &
\colhead{$\mu$} &
\colhead{$A_V$(Galactic)} &
\colhead{$A_V$(host)} &
\colhead{$M_V$} &
}
\startdata 
1954A & NGC4214 & 9.3 $\pm$ 0.2(1,2)  & 27.13 $\pm$ 0.23(3)
& 0.072 $\pm$ 0.012 & 0.05 $\pm$ 0.05(4) & $-17.95 \pm 0.31$  \\
1983N & NGC5236 & 11.3 $\pm$ 0.2(5)  & 28.25 $\pm$ 0.15$^a$(6)
& 0.228 $\pm$ 0.037 & 0.37 $\pm$ 0.37(5) & $-17.55 \pm 0.45$  \\
1984I & E323--G99 & 15.98 $\pm$ 0.20(7) & 33.66 $\pm$ 0.20$^c$(8)
& 0.344 $\pm$ 0.055 & 0.05 $\pm$ 0.05(9) & $-18.07 \pm 0.29$ \\
1984L & NGC 991 & 13.8 $\pm$ 0.2(10) & 31.85 $\pm$ 0.20$^b$
& 0.091 $\pm$ 0.015 & 0.23 $\pm$ 0.23(10) & $-18.37 \pm 0.36$ \\
1990I & NGC4650A & 15.3 $\pm$ 0.10(11) & 33.30 $\pm$ 0.28(11)  
& 0.374 $\pm$ 0.060 & 0.13 $\pm$ 0.13(11) & $-18.50 \pm 0.33$  \\   
1991D & LEDA84044 & 16.4 $\pm$ 0.3(12)  & 36.67 $\pm$ 0.05$^c$(13)
& 0.205 $\pm$ 0.033 & 0.05 $\pm$ 0.05(12) & $-20.52 \pm 0.31$ \\
1998dt & NGC 945 & 17.42 $\pm$ 0.5(14) & 34.45 $\pm$ 0.14$^c$(15)
& 0.085 $\pm$ 0.014 & 0.35 $\pm$ 0.35(4) & $-17.46 \pm 0.63$ \\
1999di & NGC 776 & 17.91 $\pm$ 0.8(14) & 34.60 $\pm$ 0.13$^c$(16)
& 0.322 $\pm$ 0.052 & 0.67 $\pm$ 0.67(4) & $-17.68 \pm 1.05$ \\
1999dn & NGC7714 & 16.48 $\pm$ 0.3(14) & 33.37 $\pm$ 0.23$^c$(16)
& 0.174 $\pm$ 0.028 & 0.05 $\pm$ 0.05(17) & $-17.11 \pm 0.38$ \\
1999ex & IC5179 & 16.63 $\pm$ 0.04(18) & 33.80 $\pm$ 0.19$^c$(8)
& 0.067 $\pm$ 0.011 & 1.39 $\pm$ 1.00(19) & $-18.63 \pm 1.02$ \\
2000H & IC 454 & 17.30 $\pm$ 0.03(20) & 34.11 $\pm$ 0.16$^c$(8)
& 0.760 $\pm$ 0.122 & 0.60 $\pm$ 0.60(4,21) & $-18.17 \pm 0.63$ \\
\enddata 
\tablenotetext{a} {Cepheid calibrated distance}
\tablenotetext{b} {Nearby Galaxies Catalog \citep{tully88}}
\tablenotetext{c} {$Luminosity\ Distance$ ($H_0$=60, $\Omega_M$=0.3,   
$\Omega_{\Lambda}$=0.7; references are for redshifts)}
\tablerefs{
(1) \cite{schaefer96}, (2) \cite{leibundgut91} and references therein,
(3) \cite{drozdovsky02}, (4) Calculated from the NaI~D line,
(5) \cite{clocc96},
(6) \cite{thim03},  
(7) Estimated from \cite{leibundgut90}, (8) NED,
(9) \cite{phillips84}, (10) \cite{wheeler85}, 
(11) \cite{elmhamdi04},  
(12) \cite{benetti02} and references therein,
(13) \cite{maza89}, (14) Estimated from \cite{matheson01},
(15) \cite{jha98}, (16) Asiago SN Catalog; \cite{barbon99},
(http://web.pd.astro.it/$\sim$supern/), (17) \cite{ayani99},
(18) \cite{stritzinger02}, (19) \cite{hamuy02},
(20) \cite{krisciunas00}, (21) \cite{benetti00}
}
\end{deluxetable} 

\clearpage 

\begin{deluxetable}{lccccccc}
\tabletypesize{\scriptsize}
\tablecaption{Absolute--Magnitude Data for SNe~Ic \label{table3}}
\tablewidth{0pt}
\tablehead{
\colhead{SN} &
\colhead{Galaxy} &
\colhead{V} &
\colhead{$\mu$} &
\colhead{$A_V$(Galactic)} &
\colhead{$A_V$(host)} &
\colhead{$M_V$} &
} 
\startdata
1962L & NGC1073 & 13.13 $\pm$ 0.10(1) & 31.39 $\pm$ 0.20$^b$
& 0.130 $\pm$ 0.021 & 0.80 $\pm$ 0.80(2) & $-19.19 \pm 0.83$ \\
1964L & NGC3938 & 13.6 $\pm$ 0.3(3) & 31.72 $\pm$ 0.14$^a$(4)
& 0.071 $\pm$ 0.011 & 0.56 $\pm$ 0.56(2) & $-18.75 \pm 0.65$ \\
1983I & NGC4051 & 13.6 $\pm$ 0.3(5) & 31.72 $\pm$ 0.14$^a$(4)
& 0.043 $\pm$ 0.007 & 0.93 $\pm$ 0.31(5) & $-19.09 \pm 0.45$ \\
1983V & NGC1365 & 13.80 $\pm$ 0.20(6) & 31.27 $\pm$ 0.05$^a$(7)
& 0.068 $\pm$ 0.011 & 0.56 $\pm$ 0.22(6) & $-18.10 \pm 0.30$ \\
1987M & NGC2715 & 14.7 $\pm$ 0.3(8) & 32.03 $\pm$ 0.23$^b$
& 0.085 $\pm$ 0.014 & 1.4 $\pm$ 0.6(9) & $-18.82 \pm 0.71$ \\
1990B & NGC4568 & 15.75 $\pm$ 0.20(10) & 30.92 $\pm$ 0.05$^a$(7)
& 0.108 $\pm$ 0.017 & 2.63 $\pm$ 1.00(10) & $-17.91 \pm 1.02$ \\
1991N & NGC3310 & 13.9 $\pm$ 0.3(11,12) & 31.84 $\pm$ 0.20$^b$
& 0.075 $\pm$ 0.012 & 1.0 $\pm$ 1.0(13) & $-19.02 \pm 1.06$ \\
1992ar & ANON & 19.54 $\pm$ 0.34(14) & 39.52 $\pm$ 0.01$^c$(15)
 & 0.048 $\pm$ 0.008 & 0.25 $\pm$ 0.25(16) & $-20.28 \pm 0.42$ \\
1994I & NGC5194 & 12.91 $\pm$ 0.02(17) & 29.62 $\pm$ 0.15(18)
& 0.115 $\pm$ 0.018 & 1.4 $\pm$ 0.5(19) & $-18.22 \pm 0.52$ \\
1997ef & UGC 4107 & 16.47 $\pm$ 0.10(20) & 33.90 $\pm$ 0.18$^c$(21)
 & 0.141 $\pm$ 0.022 & 0.05 $\pm$ 0.05(20) & $-17.62 \pm 0.21$ \\
1998bw & E184--G82 & 13.75 $\pm$ 0.10(22) & 33.13 $\pm$ 0.26$^c$(21)
 & 0.194 $\pm$ 0.031 & 0.05 $\pm$ 0.05(23) & $-19.62 \pm 0.28$ \\
1999cq & UGC11268 & 16.1 $\pm$ 0.6(24) & 35.64 $\pm$ 0.08$^c$(21)
 & 0.180 $\pm$ 0.029 & 0.39 $\pm$ 0.39(24) & $-20.11 \pm 0.72$ \\
2002ap & NGC 628 & 12.37 $\pm$ 0.04(25) & 30.41 $\pm$ 0.20$^b$
 & 0.161 $\pm$ 0.026 & 0.03 $\pm$ 0.03(26) & $-18.23 \pm 0.21$ \\
\enddata 
\tablenotetext{a} {Cepheid calibrated distance}
\tablenotetext{b} {Nearby Galaxies Catalog \citep{tully88}}
\tablenotetext{c} {$Luminosity\ Distance$ ($H_0$=60, $\Omega_M$=0.3,
$\Omega_{\Lambda}$=0.7; references are for redshifts)}
\tablerefs{
(1) \cite{schaefer95}, (2) Estimated from \cite{porter87},
(3) \cite{leibundgut91} and references therein,
(4) Same group as NGC3982; \cite{saha01}, (5) \cite{tsvetkov85},
(6) \cite{clocc97}, (7) \cite{freedman01}, (8) \cite{filippenko90},
(9) \cite{nomoto90}, (10) \cite{clocc01}, (11) \cite{tsvetkov94},
(12) \cite{korth91a} 
(13) \cite{grothues91}, (14) \cite{clocc00}, (15) \cite{phillips92},
(16) Estimated from \cite{clocc00}, (17) \cite{richmond96},
(18) \cite{feldmeier97}, (19) \cite{barth96},
(20) Estimated from \cite{iwamoto00}, (21) NED,
(22) \cite{galama98}, (23) \cite{nakamura01},
(24) Estimated from \cite{matheson00},
(25) \cite{galyam02} (data actually taken from
http://wise-obs.tau.ac.il/$\sim$avishay/local/2002ap/index.html),
(26) \cite{klose02} 
}
\end{deluxetable} 

\clearpage 

\begin{deluxetable}{lcccccc}
\tabletypesize{\normalsize}
\tablecaption{Mean Absolute Magnitudes for Various Data Sets
\label{table4}}
\tablewidth{0pt}
\tablehead{
\colhead{} & \multicolumn{2}{c}{Weighted} & \multicolumn{2}{c}{Unweighted} \\
\colhead{Data Set} &
\colhead{$\overline{M}_V$} &
\colhead{$\sigma$} &
\colhead{$\overline{M}_V$} &
\colhead{$\sigma$} &
\colhead{N}
} 
\startdata
All SE   & $-18.03 \pm 0.06$ & 0.89 & $-18.40 \pm 0.18$ & 0.94 & 27 \\ 
Bright SE  & $-20.08 \pm 0.18$ & 0.46 & $-20.13 \pm 0.19$ & 0.38 & 4 \\  
Normal SE  & $-17.77 \pm 0.06$ & 0.49 & $-18.10 \pm 0.13$ & 0.63 & 23 \\  
IIb only  & $-17.40 \pm 0.10$ & 0.15 & $-17.32 \pm 0.17$ & 0.29 & 3 \\  
Ib only  & $-18.37 \pm 0.12$ & 1.05 & $-18.18 \pm 0.27$ & 0.91 & 11 \\  
Ic only   & $-18.51 \pm 0.10$ & 0.86 & $-18.84 \pm 0.23$ & 0.83 & 13 \\  
Normal Ib  & $-17.98 \pm 0.13$ & 0.46 & $-17.95 \pm 0.16$ & 0.49 & 10 \\  
Bright Ic  & $-19.85 \pm 0.22$ & 0.37 & $-20.00 \pm 0.20$ & 0.34 & 3 \\  
Normal Ic  & $-18.14 \pm 0.12$ & 0.48 & $-18.49 \pm 0.17$ & 0.55 & 10 \\  
\enddata
\end{deluxetable} 

\clearpage 

\begin{deluxetable}{lcccccc}
\tabletypesize{\normalsize}
\tablecaption{Supernovae with Light Curves \label{table5}}
\tablewidth{0pt}
\tablehead{
\colhead{SN Name} &
\colhead{SN Type} &
\colhead{Reference} &
\colhead{SN Name} &
\colhead{SN Type} &
\colhead{Reference}
}
\startdata
1993J & IIb & (1),(2),(3) 
& 1962L & Ic & (15) \\  \hline
1996cb & IIb & (4) 
& 1983I & Ic & (16) \\   \hline
1954A & Ib & (5),(6),(7) 
& 1983V & Ic & (17) \\   \hline
1983N & Ib & (8) 
& 1987M & Ic & (18) \\   \hline
1984I & Ib & (9) 
& 1990B & Ic & (19) \\  \hline
1984L & Ib & (10) 
& 1991N & Ic & (20),(21),(22) \\  \hline
1990I & Ib & (11) 
& 1992ar & Ic & (23) \\  \hline
1991D & Ib & (12) 
& 1994I & Ic & (24),(25) \\  \hline
1999ex & Ib & (13) 
& 1997ef & Ic & (26) \\ \hline
2000H & Ib & (14) 
& 1998bw & Ic & (27),(28),(29) \\  \hline
& & & 2002ap & Ic & (30),(31),(32) \\  \hline
\enddata 
\tablerefs{
(1) \cite{barbon95}, (2) \cite{vandriel93}, (3) \cite{lewis94},
(4) \cite{qiu99},
(5) \cite{leibundgut91}, (6) \cite{schaefer96}, (7) \cite{wellmann55},
(8) \cite{clocc96},
(9) \cite{leibundgut90},
(10) \cite{baron93} and references therein, 
(11) \cite{elmhamdi04},  
(12) \cite{benetti02},
(13) \cite{stritzinger02},
(14) \cite{krisciunas00},
(15) \cite{bertola64}
(16) \cite{tsvetkov85},
(17) \cite{clocc97},
(18) \cite{filippenko90},
(19) \cite{clocc01},
(20) \cite{korth91a}, (21) \cite{korth91b}, (22) \cite{tsvetkov94},
(23) \cite{clocc00},
(24) \cite{clocc97}, (25) \cite{richmond96},
(26) \cite{iwamoto00},
(27) \cite{galama98}, 
(28) \cite{mckenzie99}, (29) \cite{sollerman00},
(30) \cite{foley03}, (31) \cite{pandey03}, (32) \cite{yoshii03}
}
\end{deluxetable} 

\clearpage 

\begin{deluxetable}{lcccccccc}
\tabletypesize{\normalsize}
\tablecaption{Parameters of the Best Light--Curve Fits \label{table6}}
\tablewidth{0pt}
\tablehead{
\colhead{SN name} &
\colhead{$E_k$} &
\colhead{$M_{ej}$} &
\colhead{$M_{Ni}$} &
\colhead{t$_{rise}$} &
\colhead{$\chi^2_r$} &
\colhead{$\delta$$M_V$} &
\colhead{N}  \\  
& (foe) & ($M_{\odot}$) & ($M_{\odot}$) & (days) & & (mag) &  
} 
\startdata
IIb \\ \hline
1993J  & 0.66(1) & 1.3 & 0.10 & 20 & 2.14 & 0.26 & 89 \\ \hline
1996cb & 0.22(2) & 0.9 & 0.08 & 20 & 1.30 & 0.18 & 44 \\ \hline \hline
Ib \\ \hline
1983N & 0.30(3) & 0.8 & 0.10 & 18 & 2.27E$-$2 & 0.45 & 9 \\ \hline
1984L & 2.16(2) & 4.0 & 0.92 & 27 & 2.63 & 0.91 & 17 \\ \hline  
1984L pk & 0.97(2) & 1.8 & 0.37 & 22 & 0.144 & 0.91 & 17 \\ \hline  
1990I & 0.67(2) & 1.2 & 0.18 & 20 & 0.743 & 1.17 & 32 \\ \hline 
1991D & 0.25(2) & 1.9 & $\geq$1.52 & 27 & 4.42 & 0.49 & 10 \\ \hline
1999ex & 0.30(2) & 0.9 & 0.25 & 19 & 0.967 & 1.05 & 71 \\ \hline \hline
Ic \\ \hline
1962L & 0.11(2) & 0.6 & 0.37 & 19 & 6.92E$-$3 & 0.85 & 11 \\ \hline
1983I & 0.33(3) & 0.7 & 0.23 & 17 & 0.140 & 0.46 & 18 \\ \hline
1983V & 0.99(2) & 1.3 & 0.15 & 19 & 3.25E$-$3 & 0.35 & 5 \\ \hline
1987M & 0.19(4) & 0.4 & 0.13 & 14 & 3.49E$-$3 & 0.72 & 4 \\ \hline
1990B & 0.55(2) & 0.9 & 0.14 & 18 & 0.229 & 1.08 & 26 \\ \hline
1992ar & 1.14(2) & 1.5 & 0.84 & 20 & 0.103 & 0.67 & 7 \\ \hline
1994I & 0.55(5) & 0.5 & 0.08 & 14 & 0.744 & 0.53 & 50 \\ \hline
1997ef & 3.26(6) & 3.1 & 0.16 & 23 & 0.491 & 0.23 & 22 \\ \hline
1998bw & 31.0(7) & 6.2 & 0.78 & 23 & 2.26 & 0.41 & 105 \\ \hline
2002ap & 2.72(8) & 1.7 & 0.14 & 19 & 1.50 & 0.23 & 60 \\ \hline \hline
\enddata 
\tablerefs{
For $E_k$/$M_{ej}$ relation --
(1) \cite{blinnikov98},
(2) $E_k$/$M_{ej}$ was calculated as described in text,
(3) \cite{shigeyama90},
(4) \cite{nomoto90},
(5) \cite{ybb95},
(6) \cite{iwamoto00},
(7) \cite{nakamura01}
(8) \cite{mazzali02}
}
\end{deluxetable} 

\begin{deluxetable}{llcclc}
\tabletypesize{\normalsize}
\tablecaption{Results Compared to Other Studies \label{table7}}
\tablewidth{0pt}
\tablehead{
\colhead{SN} &
\colhead{Study} &
\colhead{$E_k$} &
\colhead{$M_{ej}$} &
\colhead{Comments} &  \\ 
& & (foe) & ($M_{\odot}$)     
}
\startdata
IIb \\ \hline
1993J  & This paper & 0.66 & 1.3 &  \\ \hline
       & This paper (impose E$_k =$ 1 foe) & 1 & 1.6 &  \\ \hline
       & \cite{ybb95} & 1 & 2.6 &  \\ \hline
       & \cite{blinnikov98} & 1.2 & 2.45 &  \\ \hline \hline 
Ib \\ \hline
1983N  & This paper & 0.30 & 0.8 & No $^{56}$Ni mixing \\ \hline
       & This paper (impose E$_k =$ 1 foe) & 1 & 1.3 
& No $^{56}$Ni mixing \\ \hline
       & \cite{shigeyama90} & 1 & 2.7 & Extensive $^{56}$Ni mixing \\ \hline
1990I  & This paper & 0.67 & 1.2 &  \\ \hline   
       & This paper (impose E$_k =$ 1 foe) & 1 & 1.4 &   \\ \hline   
       & \cite{elmhamdi04} & 1 & 3.7  & late--times only \\ \hline  \hline  
Ic \\ \hline
1983I  & This paper & 0.33 & 0.7 & No $^{56}$Ni mixing \\ \hline
       & This paper (impose E$_k =$ 1 foe) & 1 & 1.1 & No $^{56}$Ni mixing \\ \hline
       & \cite{shigeyama90} & 1 & 2.1 & Extensive $^{56}$Ni mixing \\ \hline
1987M  & This paper & 0.19 & 0.4 & No $^{56}$Ni mixing \\ \hline
       & This paper (impose E$_k =$ 1 foe) & 1 & 0.8 & No $^{56}$Ni mixing \\ \hline
       & \cite{nomoto90} & 1 & 2.1 & Extensive $^{56}$Ni mixing \\ \hline
1997ef & This paper & 3.3 & 3.1 &  \\  \hline  
       & \cite{iwamoto00} & 8 & 7.6 &  \\ \hline 
1998bw & This paper & 31 & 6.2 &  \\ \hline 
       & \cite{nakamura01} & 50 & 10 & \\ \hline 
       & \cite{woosley99} & 22 & 6.5 & \\ \hline 
2002ap & This paper & 2.7 & 1.7 & \\ \hline 
       & \cite{mazzali02} & 4 -- 10 & 2.5 -- 5 & \\ \hline \hline  
\enddata
\end{deluxetable}  

\end{document}

%% file: figures.tex
\begin{figure}
\plotone{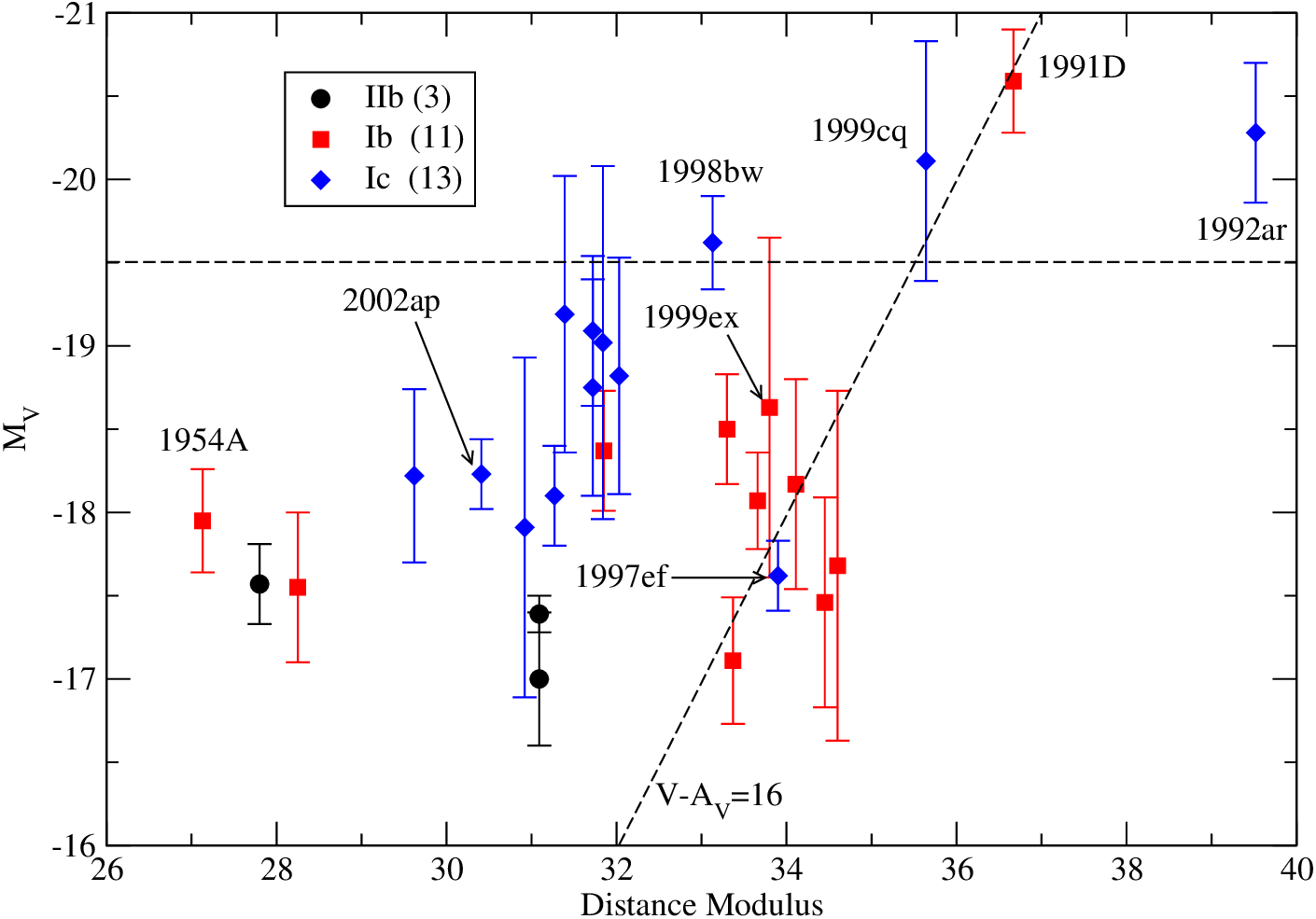}
\caption{\label{fig1} 
Absolute visual magnitude versus distance
modulus with the vertical error bars shown. Some key SNe are
labeled.
}
\end{figure} 

\clearpage 

\begin{figure}
\epsscale{0.8} 
\plotone{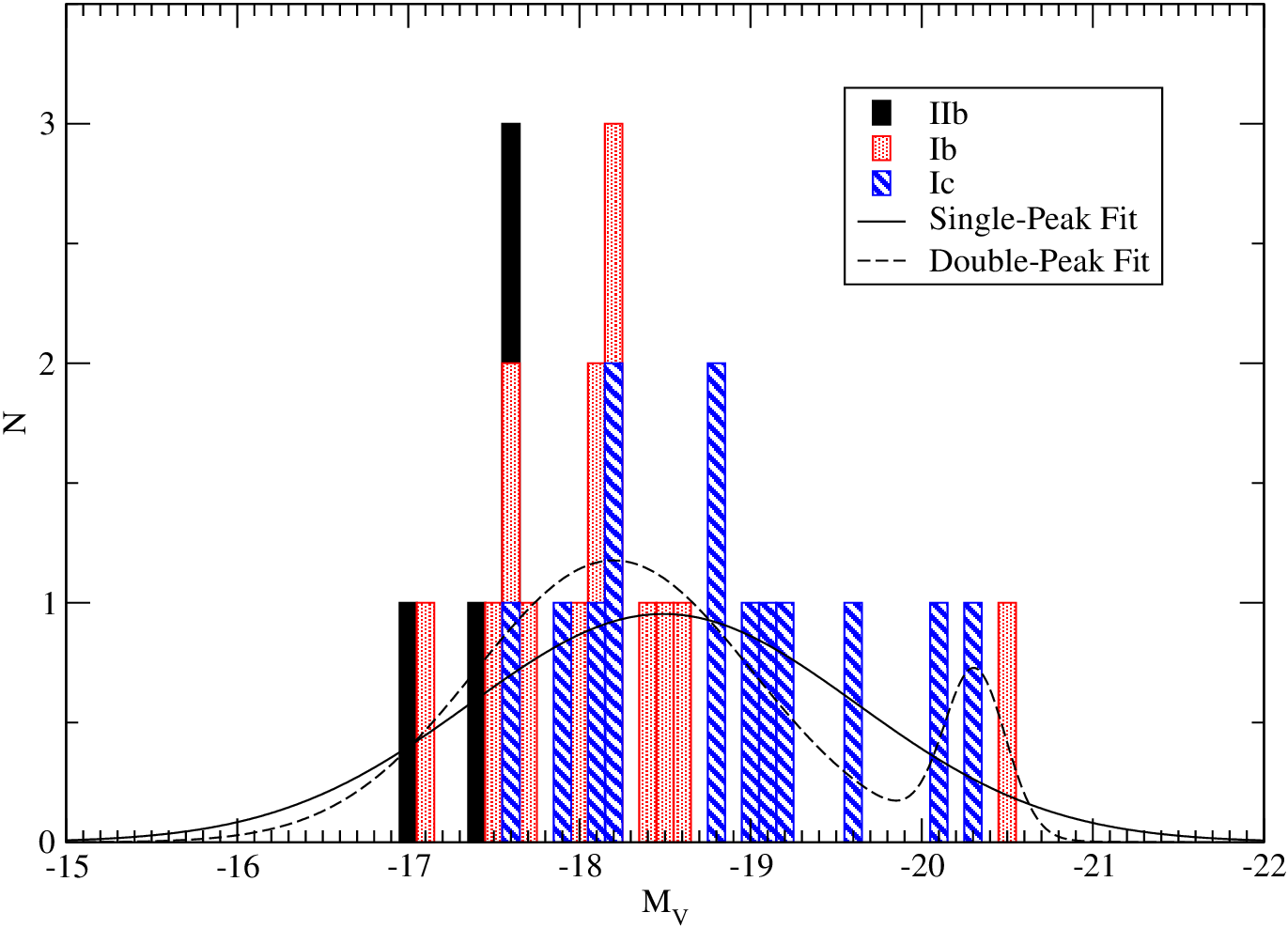}
\caption{\label{fig2} 
A histogram of SE SN absolute magnitudes, 
with 0.1 magnitude bin width. The best single--Gaussian and double--Gaussian 
fits (see text) are also shown.
}  
\end{figure} 

\clearpage 

\begin{figure}
\epsscale{0.8} 
\plotone{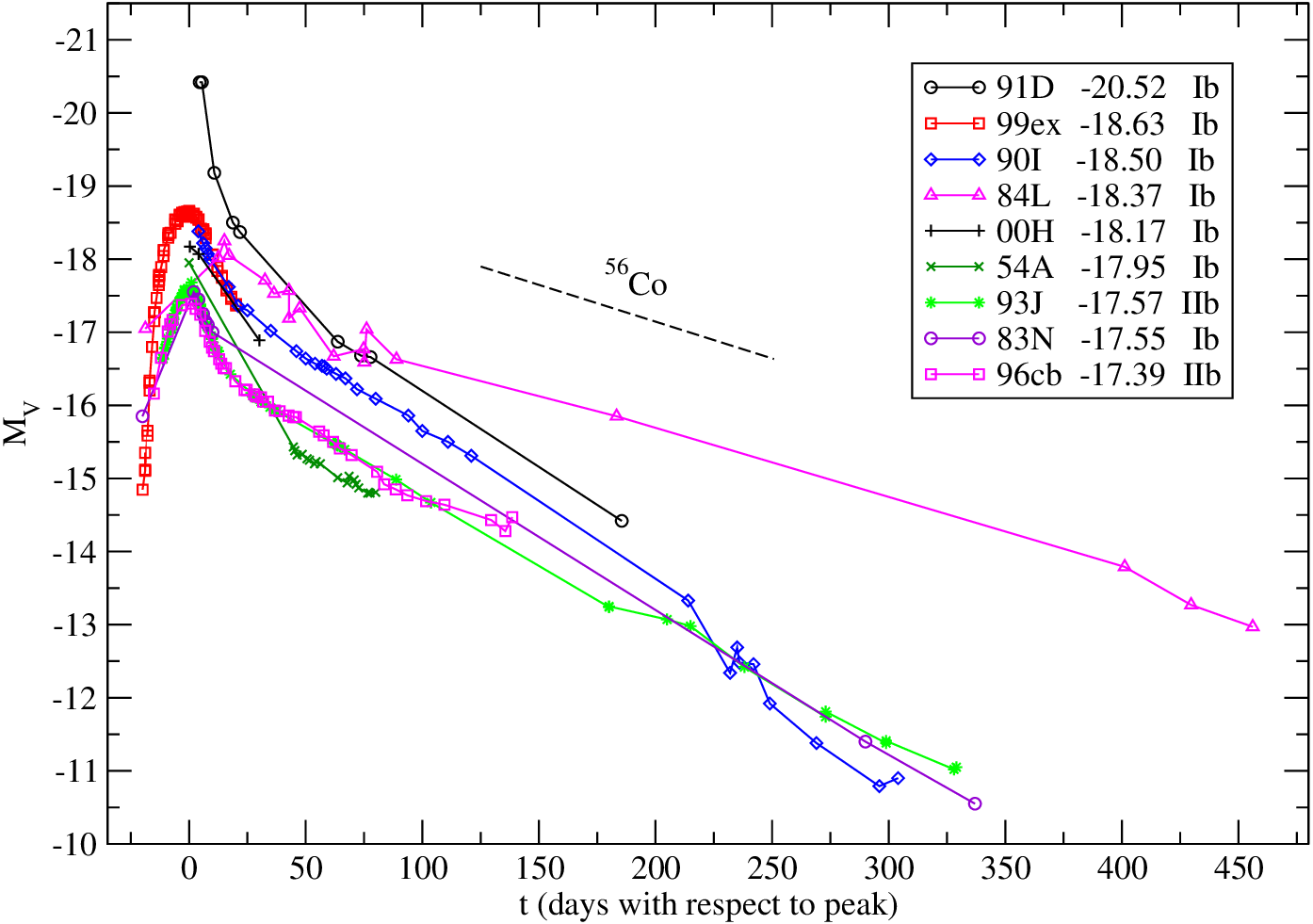}
\caption{\label{fig3} 
The absolute light curves are plotted for SNe~IIb and Ib.
The peak absolute magnitudes are given in the legend. The $^{56}$Co
decay slope is shown for reference. Solid lines are only to guide the eye.
}  
\end{figure} 

\clearpage 

\begin{figure}
\epsscale{0.8} 
\plotone{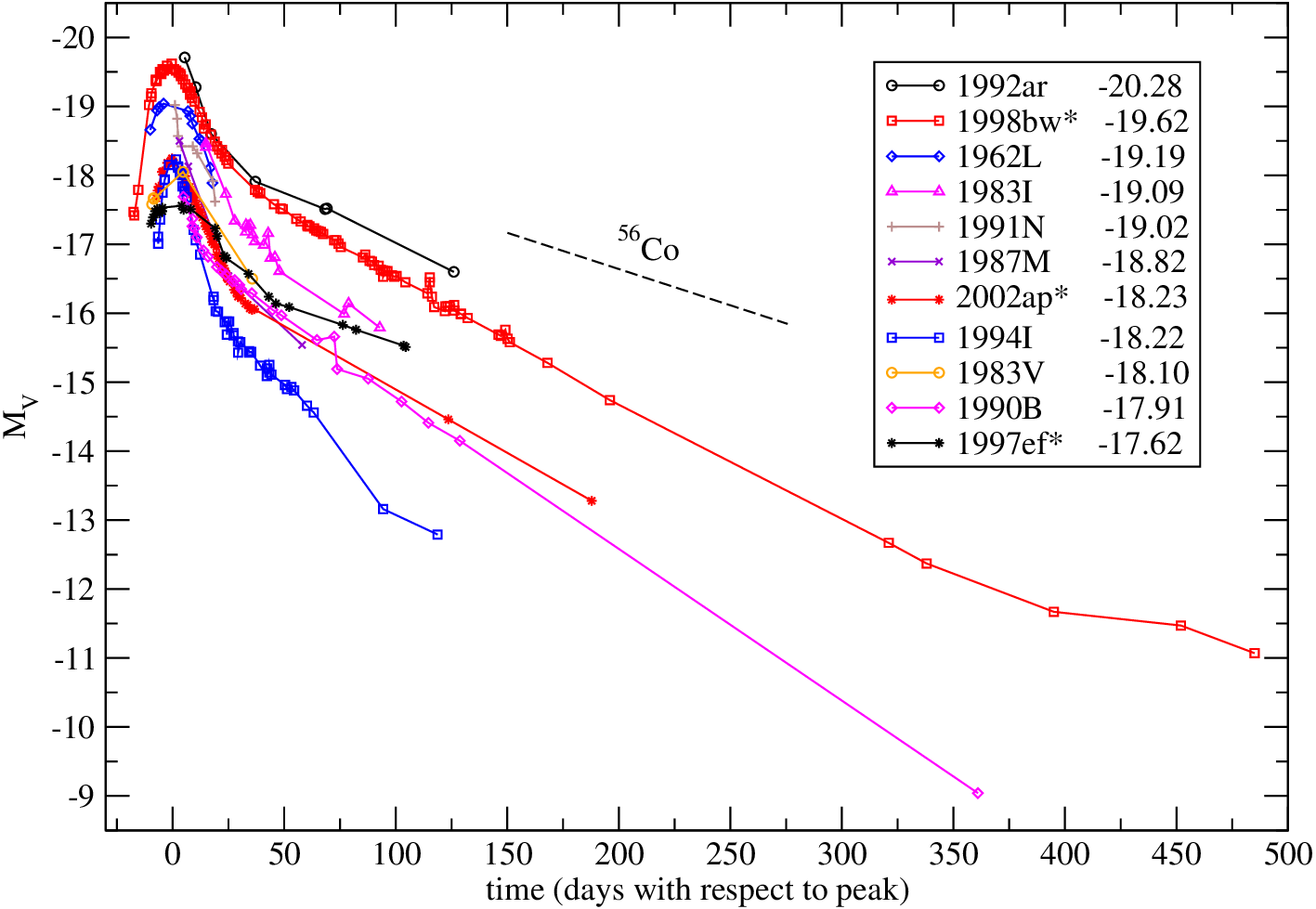}
\caption{\label{fig4} 
The absolute light curves are plotted for SNe~Ic.
The peak absolute magnitudes are given in the legend. The $^{56}$Co
decay slope is shown for reference. Solid lines are only to guide
the eye. (*=hypernovae)
}  
\end{figure} 

\clearpage 

\begin{figure}
\epsscale{0.8} 
\plotone{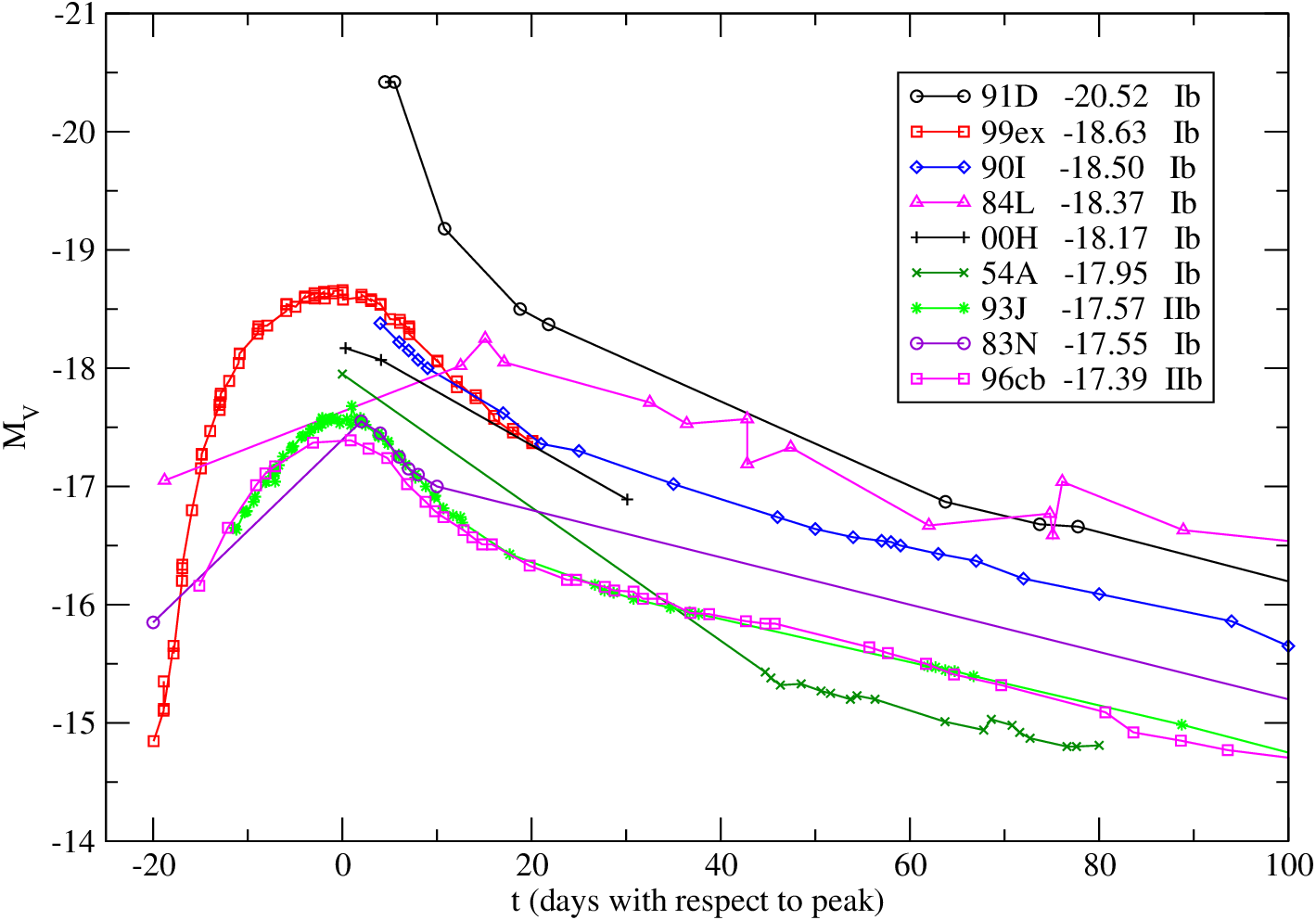}
\caption{\label{fig5} 
Same as Figure 3, except shown on a smaller time scale
around peak brightness.
}  
\end{figure} 

\clearpage 

\begin{figure}
\epsscale{0.8} 
\plotone{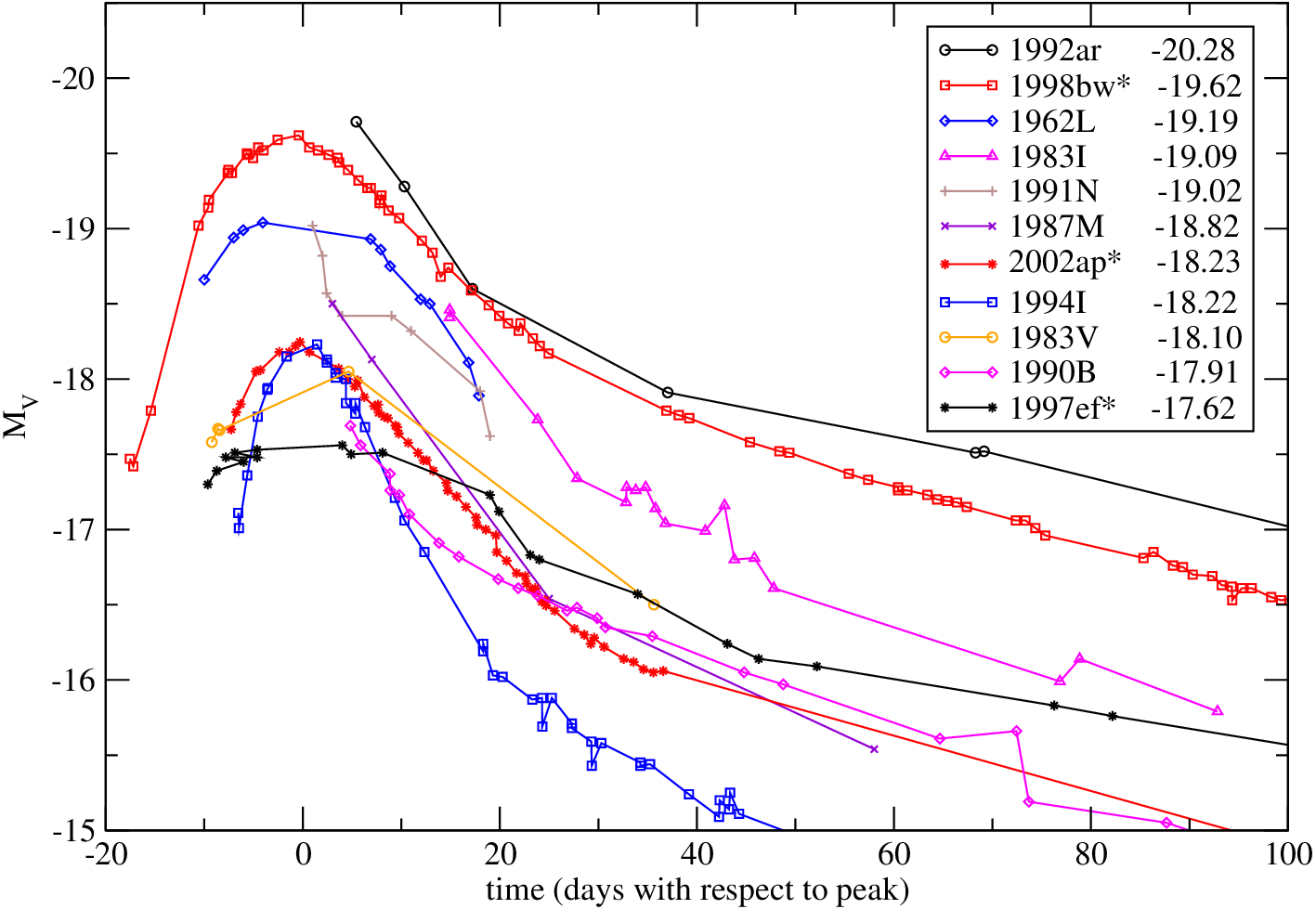}
\caption{\label{fig6} 
Same as Figure 4, except shown on a smaller time scale
around peak brightness. (*=hypernovae)
}  
\end{figure} 

\clearpage 

\begin{figure}
\epsscale{0.8} 
\plotone{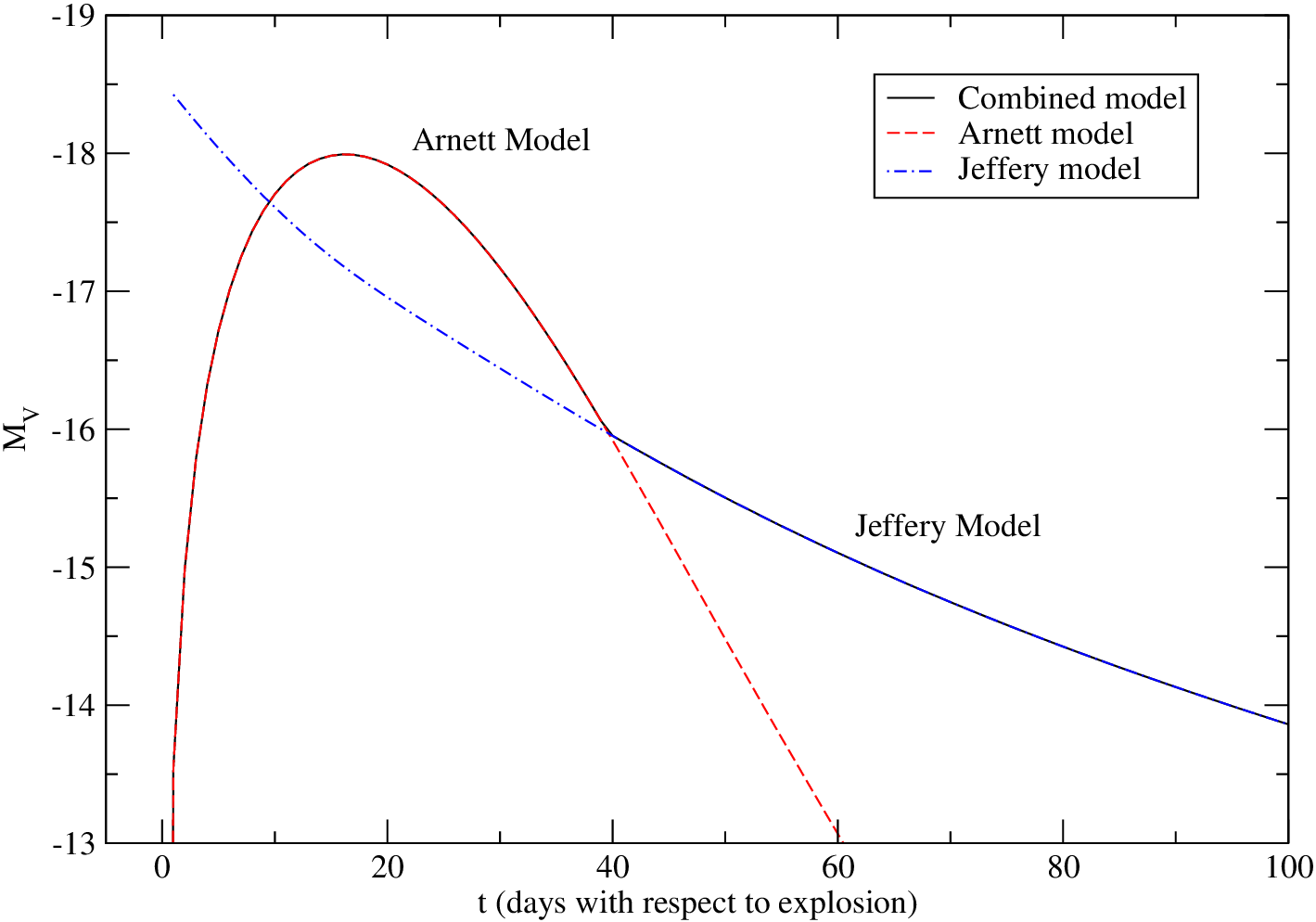}
\caption{\label{fig7} 
Here the combined model is shown by the solid line and the
Arnett and Jeffery models are shown as dashed lines.
}  
\end{figure} 

\clearpage 

\begin{figure}
\epsscale{0.8} 
\plotone{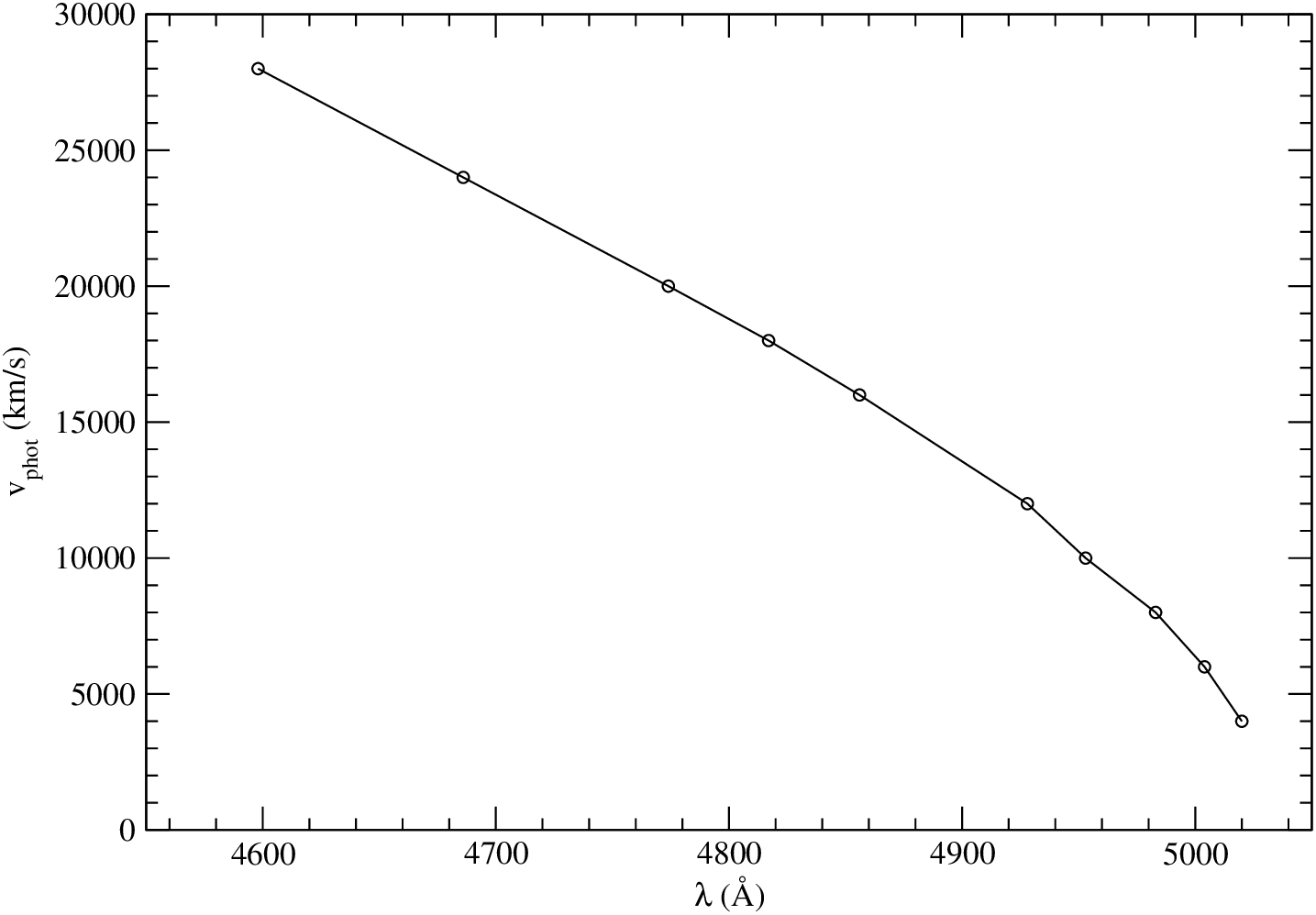}
\caption{\label{fig8} 
The relation between $v_{phot}$ and the peak of the FeII
blend near 5000\AA\ are shown as determined by {\bf Synow}.
}  
\end{figure} 

\clearpage 

\begin{figure}
\epsscale{0.8} 
\plotone{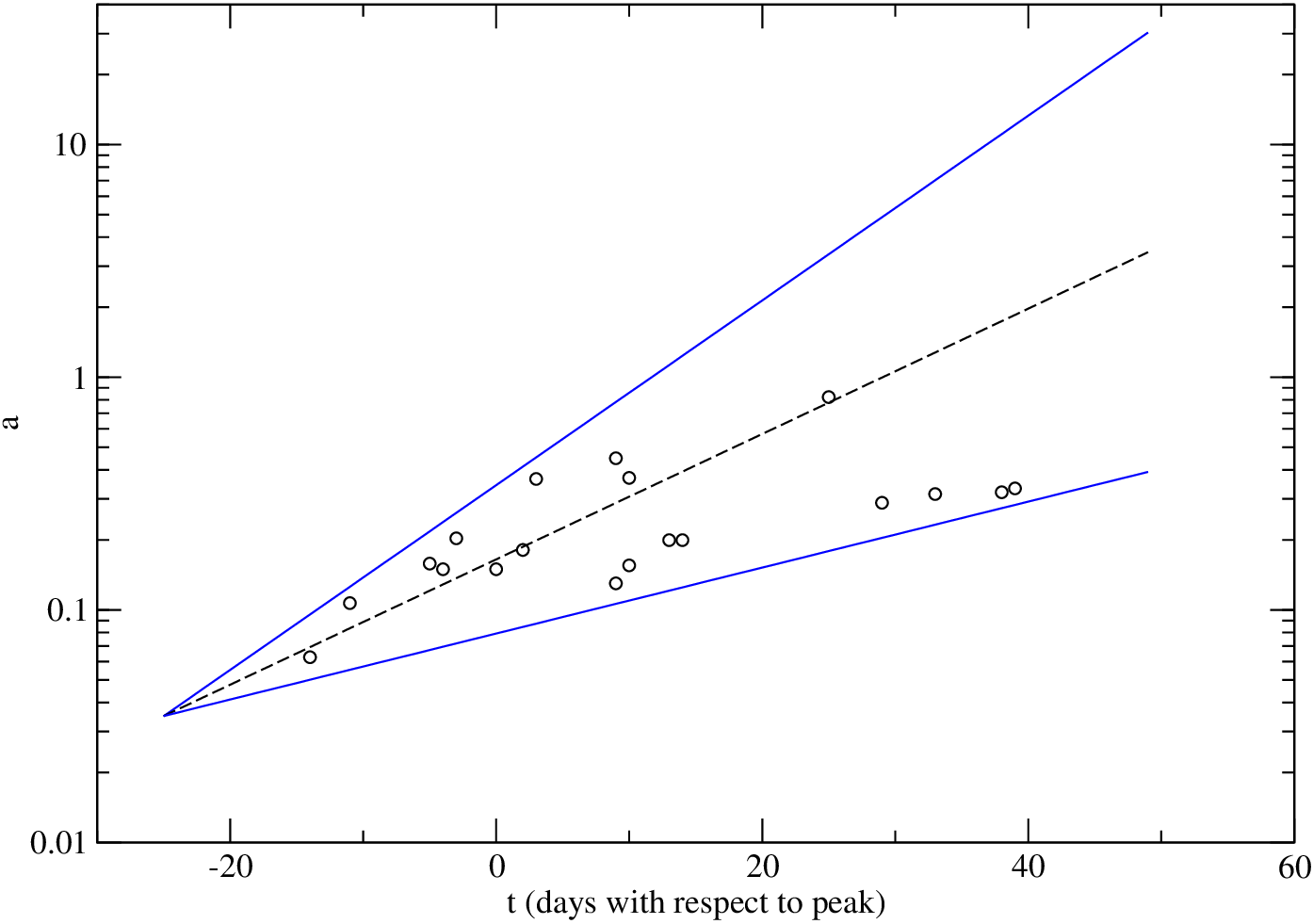}
\caption{\label{fig9} 
For normal SE~SNe, a(t) is shown on a log scale
with solid lines showing the upper and lower limits and the dashed line
showing the mean.
}  
\end{figure} 

\clearpage 

\begin{figure}
\epsscale{0.8} 
\plotone{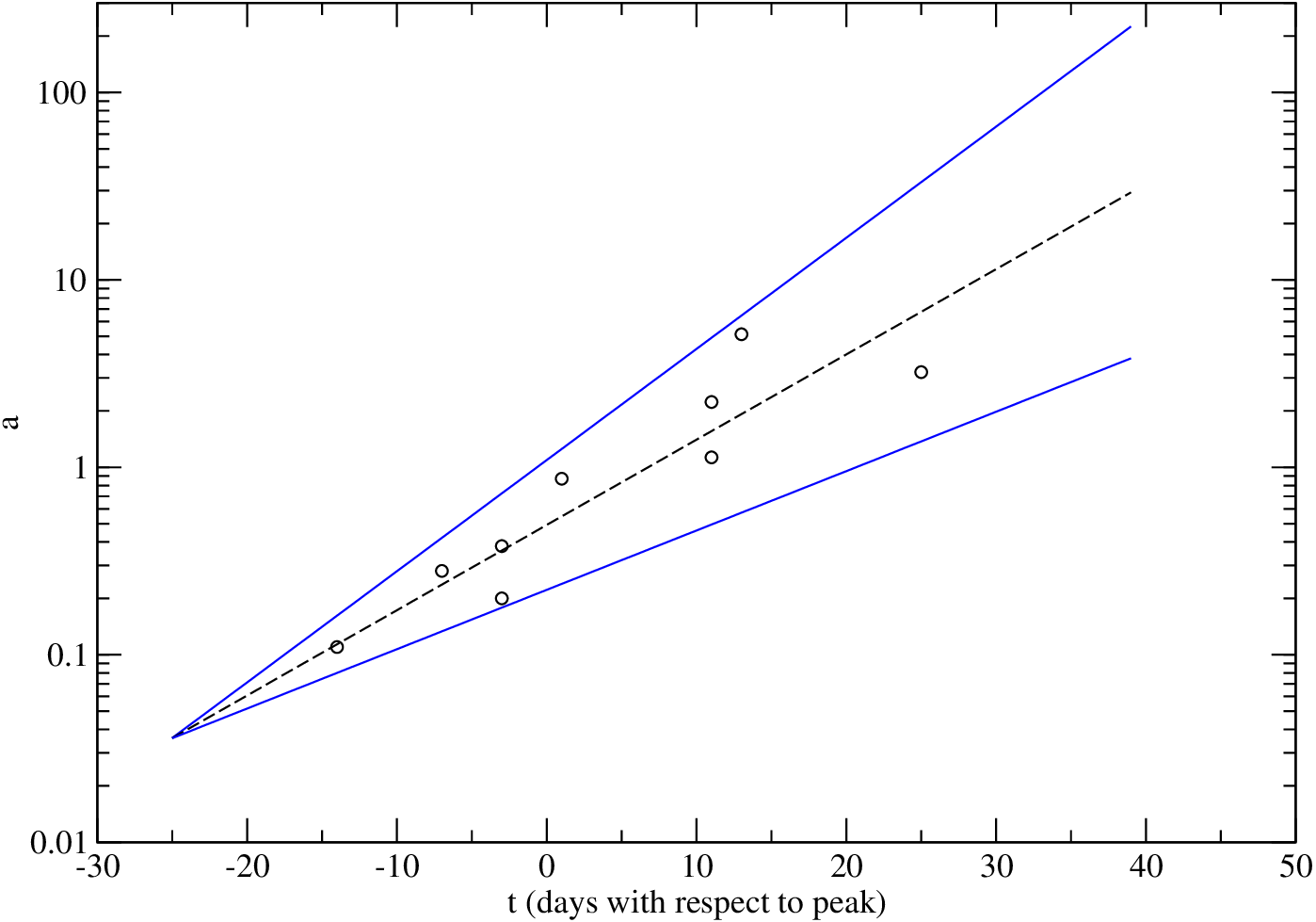}
\caption{\label{fig10} 
For hypernovae, a(t) is shown on a log scale
with solid lines showing the upper and lower limits and the dashed line
showing the mean.
}  
\end{figure} 

\clearpage 

\begin{figure}
\epsscale{0.8} 
\plotone{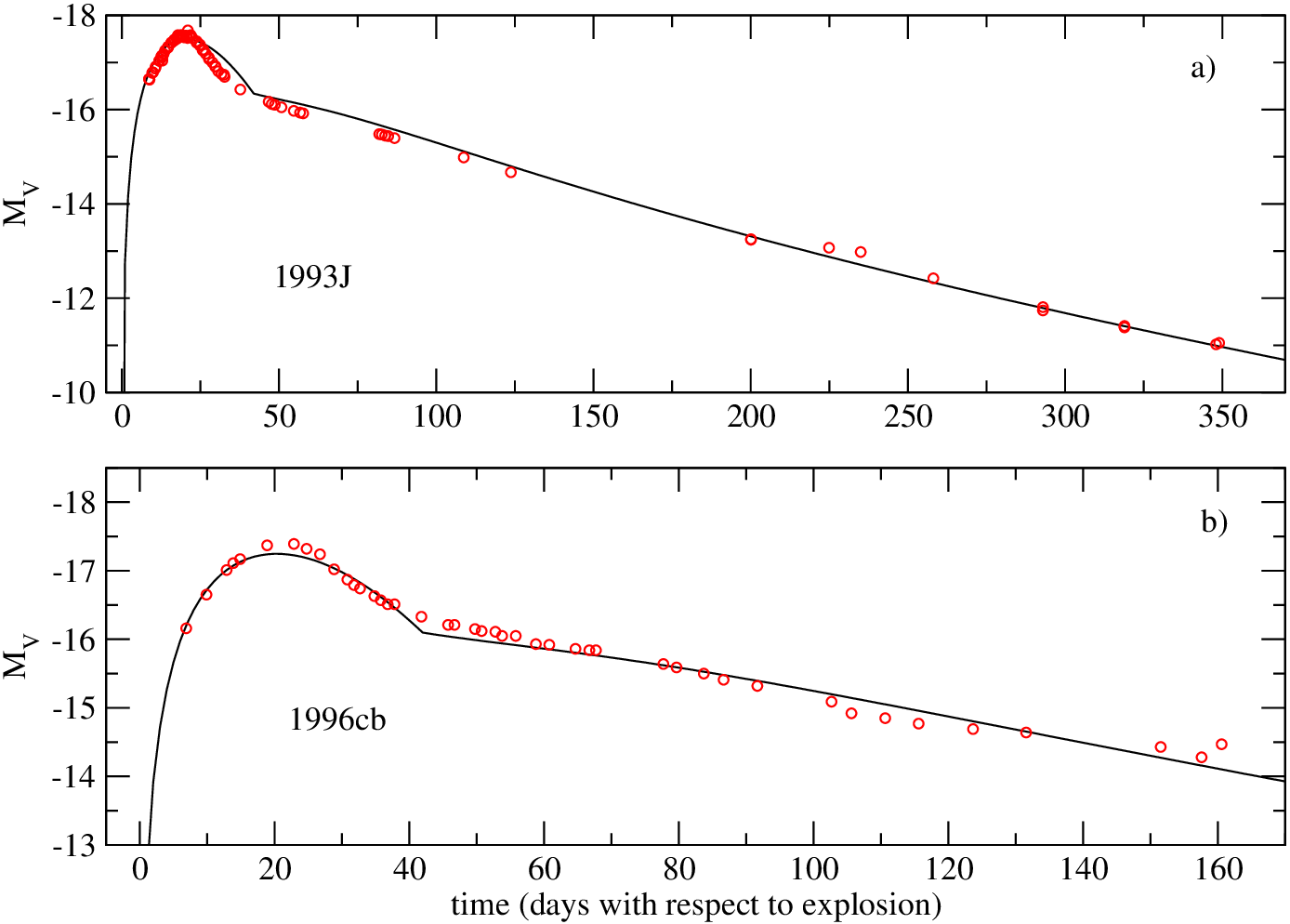}
\caption{\label{fig11} 
The best fit for the SNe~IIb (graphs are scaled independently).  
}  
\end{figure} 

\clearpage 

\begin{figure}
\epsscale{0.8} 
\plotone{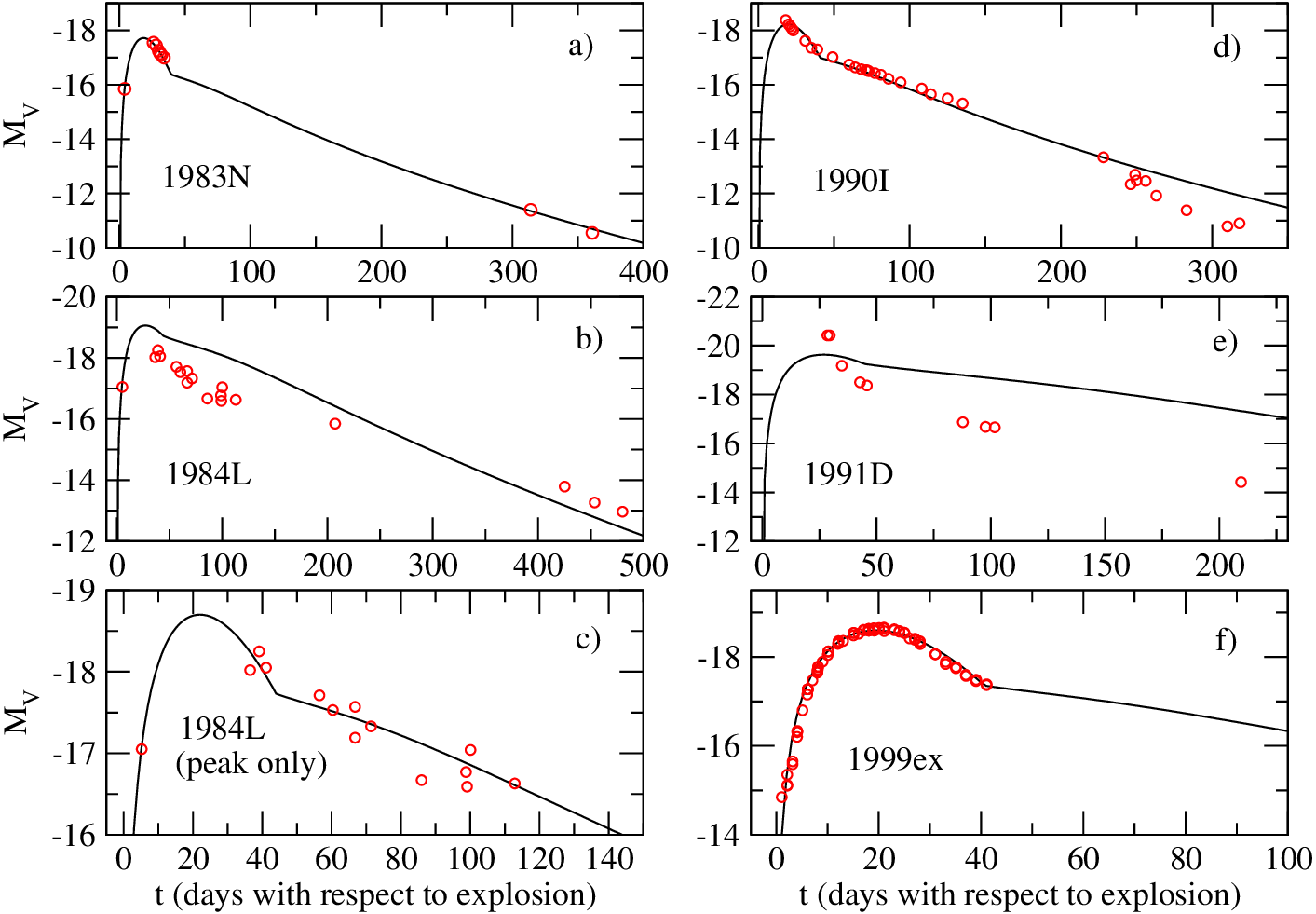}
\caption{\label{fig12} 
The best fit for the SNe~Ib (graphs are scaled independently). 
} 
\end{figure} 

\clearpage 

\begin{figure}
\epsscale{0.8} 
\plotone{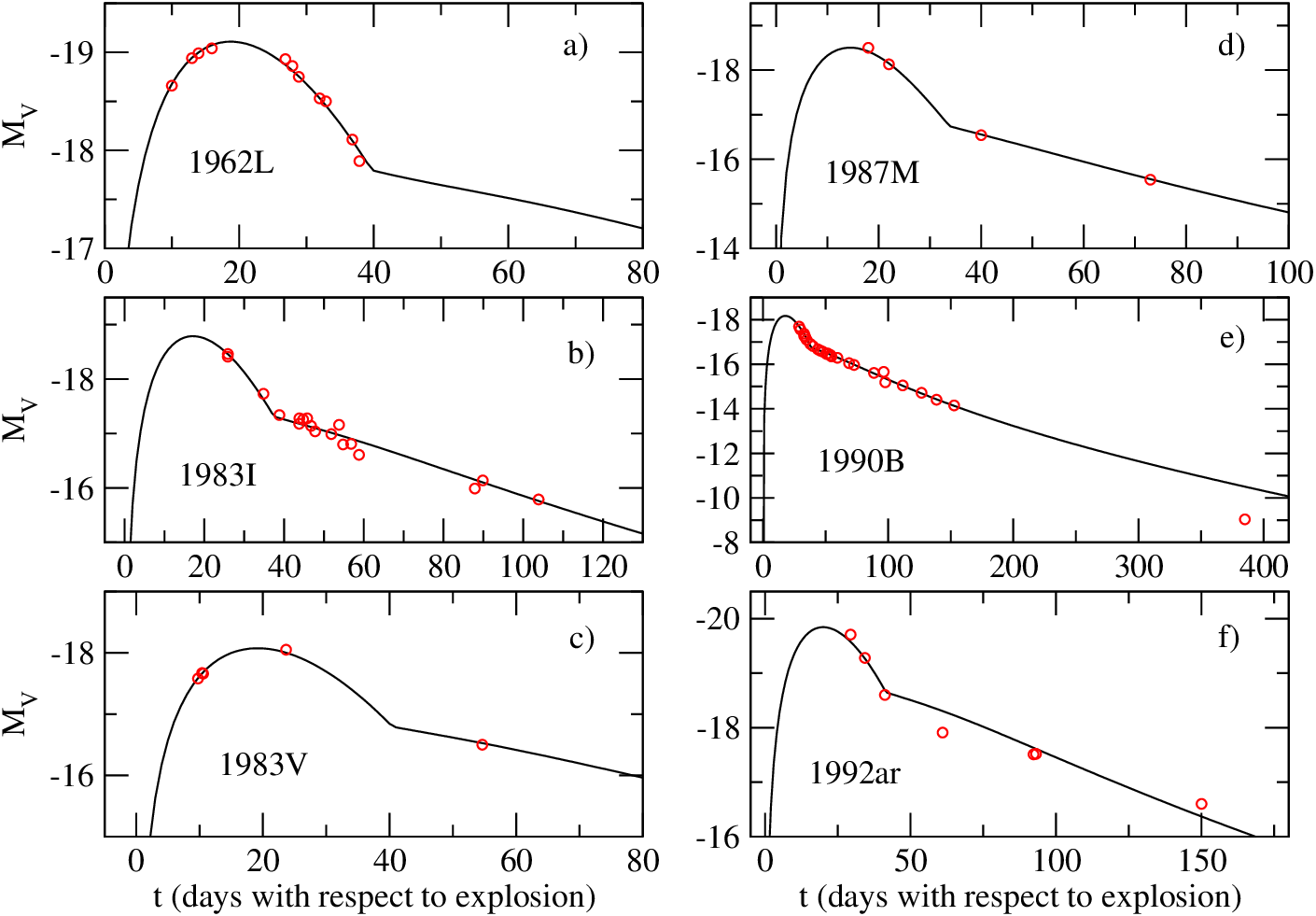}
\caption{\label{fig13} 
The best fit for SNe~Ic (graphs are scaled independently). 
}  
\end{figure} 

\clearpage

\begin{figure}
\epsscale{0.8}
\plotone{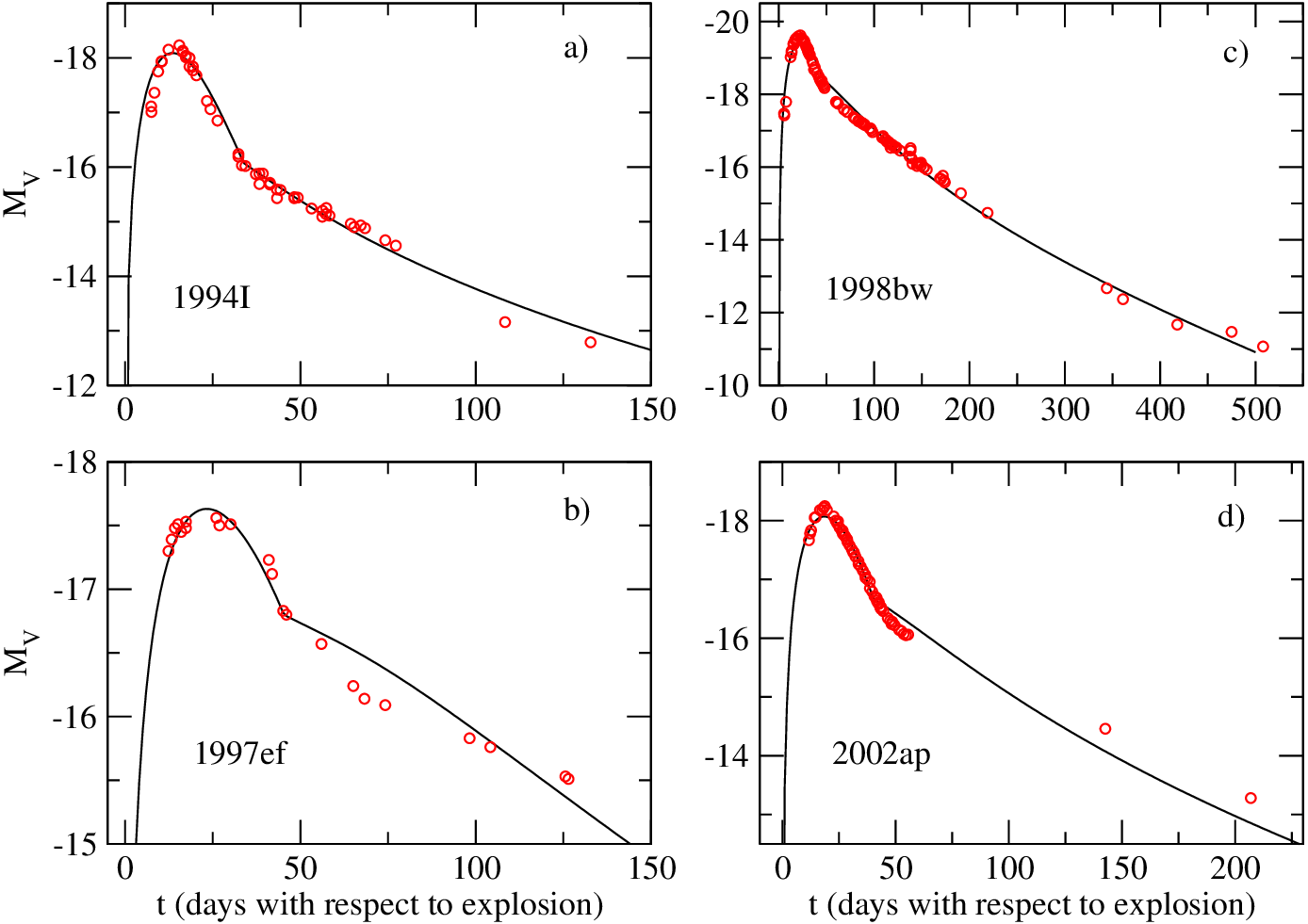}
\caption{\label{fig14}
The best fit for more SNe~Ic; b) -- c) are hypernovae (graphs are 
scaled independently).
}
\end{figure}